\documentclass[pdflatex,sn-mathphys-num]{sn-jnl}

\usepackage{graphicx}%
\usepackage{multirow}%
\usepackage{amsmath,amssymb,amsfonts}%
\usepackage{amsthm}%
\usepackage{mathrsfs}%
\usepackage[title]{appendix}%
\usepackage{xcolor}%
\usepackage{xparse}
\usepackage{textcomp}%
\usepackage{manyfoot}%
\usepackage{booktabs}%
\usepackage{algorithm}%
\usepackage{algorithmicx}%
\usepackage{algpseudocode}%
\usepackage{listings}%
\usepackage[shortlabels]{enumitem}
\usepackage{tikz-cd}
\usepackage{natbib}
\usepackage[utf8]{inputenc}
\usepackage[T1]{fontenc}
\usepackage{textcomp}
\usepackage[normalem]{ulem}
\usepackage{tikz}
\usepackage{cancel}
\usepackage{xfrac}

\DeclareMathOperator*{\argmin}{argmin}

\DeclareMathOperator*{\arcsinh}{arcsinh}

\newcommand{\bigsfrac}[2]{\scalebox{1.25}{\(\sfrac{#1}{#2}\)}}

\newcommand{\R}{\mathbb{R}}
\newcommand{\dd}{\mathrm{d}}
\newcommand{\dt}{\dd t}
\renewcommand{\t}{\mathrm{t}}

\newcommand{\reac}{\mathsf{r}}
\newcommand{\chem}{\mathsf{x}}
\newcommand{\D}{\mathrm{D}}
\newcommand{\V}{\mathsf{V}}
\renewcommand{\L}{\mathsf{L}}
\newcommand{\E}{\mathsf{E}}
\newcommand{\HH}{\mathsf{H}}
\newcommand{\e}{\mathsf{e}}
\newcommand{\w}{\mathsf{v}}
\newcommand{\om}{\omega_r(\vx)}

\newcommand{\A}{\mathcal{A}}

\newcommand{\hh}{\mathsf{h}}
\renewcommand{\tt}{\mathsf{t}}
\newcommand{\RE}{\mathbb{R}^{\E}}
\newcommand{\RV}{\mathbb{R}^{\V}}
\renewcommand{\div}{\mathrm{div}_{\HH}}
\newcommand{\grad}{\mathrm{grad}_{\HH}}

\newcommand{\EE}{\mathcal{E}}
\newcommand{\EEt}{\Tilde{\EE}}
\newcommand{\RR}{\mathcal{R}}
\newcommand{\DD}{\Sigma}
\newcommand{\DB}{\mathcal{D}}
\newcommand{\DBJ}{\mathcal{D}^{\J}}

\newcommand{\DBF}{\mathcal{D}^{\F}}
\newcommand{\DKL}{\mathcal{D}_{\textrm{KL}}}
\newcommand{\DDe}{\DD^{\textrm{ext}}}
\newcommand{\s}{\sigma}
\renewcommand{\P}{\Psi}

\newcommand{\p}{\mathfrak{p}}
\newcommand{\px}{\p_{\vx}}

\newcommand{\F}{\mathcal{F}}
\newcommand{\K}{\mathcal{K}}
\newcommand{\J}{\mathcal{J}}

\newcommand{\Fx}{\mathcal{F}_{\vx}}
\newcommand{\Fpx}{\Fx^{\textrm{b}}}
\newcommand{\Fox}{\Fx^{\textrm{c}}}
\newcommand{\Jx}{\mathcal{J}_{\vx}}
\newcommand{\Jpx}{\Jx^{\textrm{b}}}
\newcommand{\Jox}{\Jx^{\textrm{c}}}
\newcommand{\pox}{\pi_{\vx}^{\textrm{c}}}
\newcommand{\ppx}{\pi_{\vx}^{\textrm{b}}}

\newcommand{\Tj}{\T_{\vj}}
\newcommand{\Tf}{\T_{\vf}}

\newcommand{\vc}[1]{\boldsymbol{#1}}

\newcommand{\vx}{\vc{x}}
\newcommand{\vz}{\vc{c}}
\newcommand{\xss}{\vx^{\textrm{ss}}}
\newcommand{\xssd}{\dot{\vx}^{\textrm{ss}}}

\newcommand{\vt}{\vc{\theta}}
\newcommand{\vj}{\vc{j}}
\newcommand{\vjx}{\vj_{\vx}}
\newcommand{\vjc}{\vj^{\textrm{c}}}
\newcommand{\vjb}{\vj^{\textrm{b}}}

\newcommand{\vfc}{\vf^{\textrm{c}}}
\newcommand{\vfb}{\vf^{\textrm{b}}}
\newcommand{\vv}{\vc{v}}
\newcommand{\vf}{\vc{f}}

\newcommand{\vm}{\vc{\mu}}

\newcommand{\fe}{\vf^{\textrm{ext}}}
\newcommand{\fet}{\overline{\vf}^{\textrm{ext}}}
\newcommand{\fex}{\vf^{\textrm{ext}}_{\vx}}
\newcommand{\fext}{\overline{\vf}^{\textrm{ext}}_{\vx}}
\newcommand{\se}{\s^{\textrm{ext}}}

\renewcommand{\k}{\kappa}
\newcommand{\vk}{\vc{k}}
\newcommand{\vkk}{\vc{\k}}

\newcommand{\gJ}{\mathsf{g}^{\mathcal{J}}}

\newcommand{\gJxj}{\gJ_{\vx,\vj}}

\newcommand{\gF}{\mathsf{g}^{\mathcal{F}}}
\newcommand{\gFxf}{\gF_{\vx,\vf}}

\newcommand{\T}{\mathcal{T}}
\newcommand{\X}{\mathbf{X}}
\newcommand{\img}{\mathrm{img}}
\renewcommand{\ker}{\mathrm{ker}}

\NewDocumentCommand{\com}{+m}{{\color{black}#1}}

\theoremstyle{thmstyleone}%
\newtheorem{theorem}{Theorem}[section]
\newtheorem*{theorem*}{Theorem}
\newtheorem*{definition*}{Definition}
\newtheorem{definition}{Definition}[section]
\newtheorem{lemma}[theorem]{Lemma}%

\theoremstyle{thmstyletwo}%
\newtheorem{example}{Example}[section]%
\newtheorem{remark}{Remark}[section]
\newtheorem*{remark*}{Remark.}%

\theoremstyle{thmstylethree}%

\numberwithin{equation}{section}

\begin{document}

\title[Information geometry of gradient flow systems]{Information geometry of perturbed gradient flow systems on hypergraphs: A perspective towards nonequilibrium physics}


\author*[1,2]{\fnm{Dimitri} \sur{Loutchko}}\email{dimitri@sat.t.u-tokyo.ac.jp}

\author[3]{\fnm{Keisuke} \sur{Sugie}}

\author*[1,3,4]{\fnm{Tetsuya} \fnm{J} \sur{Kobayashi}} \email{tetsuya@mail.crmind.net}

\affil*[1]{\orgdiv{Institute of Industrial Science}, \orgname{The University of Tokyo}, \orgaddress{\street{4-6-1, Komaba}, \city{Meguro-ku}, \postcode{153-8505}, \state{Tokyo}, \country{Japan}}}

\affil[2]{\orgdiv{Center for Interdisciplinary AI and Data Science}, \orgname{Ochanomizu University}, \orgaddress{\street{2-1-1 Otsuka}, \city{Bunkyo-ku}, \postcode{112-8610}, \state{Tokyo}, \country{Japan}}}

\affil[3]{\orgdiv{Department of Mathematical Informatics, Graduate School of Information Science and Technology}, \orgname{The University of Tokyo}, \orgaddress{\street{7-3-1 Hongo}, \city{Bunkyo-ku}, \postcode{113-8656}, \state{Tokyo}, \country{Japan}}}

\affil[4]{\orgdiv{Universal Biology Institute}, \orgname{The University of Tokyo}, \orgaddress{\street{7-3-1 Hongo}, \city{Bunkyo-ku}, \postcode{113-8654}, \state{Tokyo}, \country{Japan}}}


\abstract{
This article concisely reviews the link between gradient flow systems on hypergraphs and information geometry, which has been established within the last five years.
Gradient flow systems describe a wealth of physical phenomena and provide powerful analytical techniques which are based on the variational energy-dissipation principle.
Modern nonequilibrium physics has complemented this classical variational structure with thermodynamic uncertainty relations, speed limits, entropy production rate decompositions, and many more.
In this article, we formulate these modern principles within the framework of perturbed gradient flow systems on hypergraphs.
In particular, we discuss the geometry induced by the Bregman divergence, the physical implications of dual foliations, as well as the corresponding infinitesimal Riemannian geometry for gradient flow systems.
Through the geometric perspective, we are naturally led to new concepts such as moduli spaces for perturbed gradient flow systems and thermodynamic area, which is crucial for understanding speed limits.
We hope to encourage the readers working in either of the two fields to further expand on and foster the interaction between the two fields.
}




\maketitle

\section{Introduction}\label{sec:intro}

The gradient flow formalism provides a wealth of analytical techniques and physical insight to study evolution equations of the type
\begin{align} \label{eq:gf_intro}
      \dot{\vx} = \D_{\vm} \RR^*(\vx, -\D_{\vx} \EE(\vx)),
\end{align}
where $\EE: \X \rightarrow \R$ is a driving functional such as the system energy or entropy, and $\RR^*: \T^*\X \rightarrow \R$ is the dissipation potential.

If a system has a gradient flow structure, it allows to deal with stability and convergence properties \cite{otto2000generalization,bakry2013analysis}, to construct weak solutions \cite{mielke2002variational,maso2006quasistatic,ambrosio2005gradient,mielke2015rate,peletier2022jump},
and to implement various limiting procedures while retaining thermodynamic consistency \cite{peletier2022gamma, disser2015gradient,gladbach2020scaling,mielke2021edp,mielke2023non}.
Applications include classical reaction\com{-}diffusion systems \cite{mielke2011gradient}, the Boltzmann equation \cite{erbar2023gradient}, Cahn-Hillard equations \cite{garcke2020weak}, the physics of thin layers \cite{glitzky2024drift}, diodes \cite{kirch2021electrothermal}\com{,} and semiconductor devices \cite{sawatzki2018balance}, among others.
We refer to the article \cite{PeletierSchlichting2023} for an introduction and an \com{extensive} review of gradient flow systems.

In a generalized sense, the system (\ref{eq:gf_intro}) describes the evolution of an \com{equilibrium} system as it \com{minimizes} the driving functional $\EE$.
In this article, we deal with perturbed gradient flow systems on hypergraphs \cite{mielke2016deriving,renger2018,kobayashi2024information} as they can \com{accommodate} an external driving of the system via an external force term $\fex$ in the continuity equation
\begin{align}
\begin{split}
\label{eq:dynamics_HG_1_perturbed_intro}
    \dot{\vx} &= -\div \vj, \\ 
    \vj &= \D_{\vf} \P^*(\vx, \grad[-\D_{\vx} \EE(\vx)] + \fex)
\end{split}
\end{align}
and thus can describe genuine nonequilibrium situations.
Thereby, $\div$ and $\grad$ denote discrete divergence and gradient operators associated to a hypergraph, and $\P^*:\F \rightarrow \R$ is the dissipation potential on the space of thermodynamic forces. 
In particular, jump processes and chemical reaction networks (CRNs) have been widely investigated within the gradient flow formalism \cite{adams2013large,mielke2014,mielke2017,mielke2021edp,renger2018flux}.

In recent decades, in nonequilibrium and statistical physics, and especially within the field of stochastic thermodynamics \cite{pigolotti2021stochastic,seifert2025stochastic,falasco2025macroscopic}, the concepts of free energy minimization in equilibrium systems have been extended to the nonequilibrium situation, and have been complemented with newly discovered principles which include, among others, fluctuation theorems \cite{seifert2012stochastic,korbel2021stochastic,martins2025brief}, thermodynamic uncertainty relations \cite{barato2015thermodynamic,horowitz2020thermodynamic,falasco2020unifying,koyuk2022thermodynamic,kamijima2023thermodynamic,ray2023thermodynamic}, speed limits \cite{vo2020unified,yoshimura2021thermodynamic,nagayama2025infinite}, and entropy production rate decompositions \cite{dechant2022geometric,kobayashi2022geometry,yoshimura2023housekeeping}.

Starting with the pioneering work \cite{ito2018stochastic}, the importance and \com{utility} of information geometry to concisely derive and analyze such principles for Markov jump processes and chemical reaction networks has become ever more apparent and successful \cite{ito2020unified,ito2020stochastic,yoshimura2021information,sughiyama2021,kobayashi2021,kobayashi2022geometry,loutchko2022,sughiyama2022chemical,ito2022information,ito2024geometric,mizohata2024information,kobayashi2024information}.
In \cite{kobayashi2024information}, it was noted that, to a great extent, such results do not depend on the specifics of physical system but all that is mathematically required in the derivations is the existence of a gradient flow structure (\ref{eq:dynamics_HG_1_perturbed_intro}).
As an application, information\com{-}geometric \com{entropy production rate} decompositions have been treated from this general point of view \cite{kobayashi2022geometry,kolchinsky2022information}.
Moreover, within the statistics community, works which combine the two fields have started to emerge \cite{li2022transport,ay2024information}.

In this article, we continue and expand on building the mathematical framework of modern \com{nonequilibrium} physics based on the information geometry of gradient flow systems.
Following \cite{kobayashi2024information}, we introduce the information geometry of gradient flow systems and \com{demonstrate} how it can be used to concisely derive the above\com{-}mentioned modern physical principles purely from the existence of a gradient flow structure.

In Section \ref{sec:gradient_flow_basics}, we introduce and discuss the basics of gradient flow systems, then extend them to gradient flow systems on hypergraphs in Section \ref{sec:gradient_flow_hypergraphs}, and add the perturbation in Section \ref{sec:perturbed_systems}. Thereby we focus on the energy-dissipation principle and how it is influenced by the perturbation.

In Section \ref{sec:Bregman_LDT}, we develop the corresponding information geometry, and reformulate the energy-dissipation principle by using the Bregman divergence.
We also discuss the large deviations theoretic interpretation of this quantity.
In Section \ref{sec:Riemannian_IG}, this stochastic interpretation is used to relate the Fisher metric from information geometry to the covariance matrix on the flux space.
We conclude this section by deriving the dual \com{orthogonal foliations} of the flux and force spaces, discuss their physical significance, and derive a new theorem on the moduli spaces of perturbed gradient flow systems on a hypergraph in Section \ref{sec:dual_foliations}.

Section \ref{sec:noneq_physics} is the centerpiece of this article as we provide concise formulations of modern results in \com{nonequilibrium} physics through the setup from Section \ref{sec:Info_Gradient_flow}.
In particular, we do not assume any specific model or structure of the system but only require it to be a gradient flow system of the form (\ref{eq:dynamics_HG_1_perturbed_intro}).
In Section \ref{sec:decomposition} we present the information\com{-}geometric dissipation rate decomposition from \cite{kobayashi2022geometry}.
It is worth noting that this decomposition does not require the force-flux pair to be conjugate and thus is valid for fluctuating systems.
In Section \ref{sec:Riemannian_application}, we present a package of techniques to obtain \com{thermodynamic uncertainty relations}, thermodynamic speed limits\com{,} and another information\com{-}geometric dissipation rate decomposition which is based on the Riemannian \com{geometric} aspects of the theory.

Finally, we discuss future perspectives in Section \ref{sec:discussion}.

\section{Gradient flow systems} \label{sec:gradient_flow}

In this section, we introduce and discuss the formalism of gradient flow systems.
We follow the review article \cite{PeletierSchlichting2023} \com{but add more differential-geometric flavor to the exposition in preparation for Sections \ref{sec:Info_Gradient_flow} and \ref{sec:noneq_physics}.
As we do not require advanced differential-geometric manipulations, we use coordinate-free notation to keep the exposition concise.
For readers unfamiliar with this notation, we provide a more detailed exposition in Appendix \ref{app:diff_geom}.
}

\subsection{Basics of gradient flow systems} \label{sec:gradient_flow_basics}

A gradient flow system is given by an evolution equation with an additional variational structure.
In its basic form, the evolution equation of the system with the state space $\X$ (\com{in most applications, the elements $\vx \in \X$ are positive or probability measures over an underlying discrete or continuous space}) is given by
\begin{align} \label{eq:dynamics}
      \dot{\vx} = \D_{\vm} \RR^*(\vx, -\D_{\vx} \EE(\vx)).
\end{align}
Here, $\EE: \X \rightarrow \R$ is the {\it driving functional} such as the system energy or entropy, and $\RR^*: \T^*\X \rightarrow \R$ is the {\it dissipation potential} on the cotangent bundle with coordinates $(\vx, \vm) \in \T^*\X$.
The operators $\D_{\vx}$ and $\D_{\vm}$ denote derivatives with respect to $\vx$ and $\vm$.
This gradient system is denoted by the triple $(\X, \RR^*, \EE)$.

This structure has a clear physical meaning:
The term $-\D_{\vx} \EE(\vx)$ corresponds to a thermodynamic force, and the operation $\D_{\vm} \RR^*(\vx, .)$ converts this force to a velocity.
This is a generalization of the Onsager relations \cite{onsager1931reciprocal}, which constitute a linear relation between forces and velocities, and are recovered when $\RR^*(\vx, \vm)$ is quadratic in the second argument \cite{mielke2003energetic,mielke2005evolution,mielke2011gradient,mielke2015rate}.
Analogous to the Onsager relations, $\RR^*(\vx, \vm)$ and its Legendre transform in the second argument $\RR(\vx, \vv)$ are required to satisfy
\begin{enumerate}[(a), leftmargin=2cm]
    \item $\RR^*(\vx, .)$ and $\RR(\vx, .)$ are strictly convex in the second argument,
\label{prop:10}
    \item $\RR^*(\vx, .)$ and $\RR(\vx, .)$ are 1-coercive\footnote{
\com{1-coercivity means that the functions grow superlinearly at infinity, i.e, 
\begin{equation}
    \frac{\RR(\vx, \vv)}{\| \vv\|} \rightarrow \infty \textrm{  as  } \| \vv \| \rightarrow \infty,
\end{equation}
and analogously for $\RR^*(\vx, .)$.
This is a standard requirement for a gradient flow system as it ensures that the Legendre duality between $\RR$ and $\RR^*$ is finite-valued and that the Legendre duality induces an isomorphism of vector spaces between $\T_{\vx}\X$ and $\T^*_{\vx}\X$.
}},
    \item $\RR^*(\vx, .)$ and $\RR(\vx, .)$ are even in the second argument,
\label{prop:20}
    \item $\min \RR^*(\vx, .) = \min \RR(\vx, .) = 0$, and $\RR^*(\vx, \vc{0}) = \RR(\vx, \vc{0}) = 0$
\label{prop:30}
\end{enumerate}
for all $\vx \in \X$.
The Young-Fenchel relation reads
\begin{align} \label{eq:Young-Fenchel}
    \RR(\vx,\vv) + \RR^*(\vx,\vm) \geq \langle \vv, \vm \rangle
\end{align}
with equality iff $\vv = \D_{\mu} \RR^*(\vx,\vm)$ and $\vm = \D_{\vv} \RR(\vx,\vv)$ hold.
A central quantity of a gradient flow system is the {\it dissipation rate}\footnote{
\label{footnote:dissipation}
The dissipation rate $\s$ is often referred to as the entropy production rate in the physics literature.
In particular, the concepts discussed in Section \ref{sec:noneq_physics} are usually stated with $\s$ being identified with the \com{entropy production rate}.
However, the dissipation rate has a more general meaning in the context of gradient flow systems as the driving functional $\EE$ can be any thermodynamic potential and thus $\s$ can describe the dissipation of various thermodynamic potentials, depending on the setup.}
$\s$ at a point $(\vx,\vv) \in \T\X$, defined as 
\begin{align}
    \s(\vx,\vv) := \RR(\vx,\vv) + \RR^*(\vx,-\D_{\vx} \EE(\vx)).
\end{align}
The dissipation rate is lower bounded by $\langle \vv, -\D_{\vx} \EE(\vx) \rangle$ according to (\ref{eq:Young-Fenchel}), and if $\vv = \dot{\vx}$ is the gradient flow vector field, then equality holds and the dissipation rate becomes $\s(\vx,\dot{\vx}) = \langle \dot{\vx}, -\D_{\vx} \EE(\vx) \rangle = - \dot{\EE}(\vx)$.
Together with the non-negativity of the dissipation functions, this implies that $\dot{\EE}(\vx) = - [\RR(\vx,\dot{\vx}) + \RR^*(\vx,-\D_{\vx} \EE(\vx))] \leq 0$ with equality iff $\dot{\vx} = \vc{0}$ and $-\D_{\vx} \EE(\vx) = \vc{0}$ \com{hold;} thus\com{,} $\EE$ is a Lyapunov function for the dynamics.

Globally, the {\it total dissipation functional} $\DD[\cdot]$ is defined for a differentiable curve $\vx : [0,T] \rightarrow \X$ as
\begin{align} \label{eq:dissipation}
    \DD[\vx(\cdot)] := \int_0^T \RR(\vx,\dot{\vx}) + \RR^*(\vx,-\D_{\vx} \EE(\vx)) \dt.
\end{align}
Following the discussion of the dissipation rate, one verifies the energy-dissipation relation
\begin{align} \label{eq:EDP}
    \EE(\vx(T)) + \int_0^T \RR(\vx,\dot{\vx}) + \RR^*(\vx,-\D_{\vx} \EE(\vx)) \dt \geq \EE(\vx(0))
\end{align}
with the equality if and only if $\dot{\vx}$ and $-\D_{\vx} \EE(\vx)$ are Legendre dual, i.e., if and only if the gradient flow equation (\ref{eq:dynamics}) holds true.
This equivalence is known as De Giorgi's energy-dissipation principle (EDP).

\subsection{Gradient flow systems on hypergraphs} \label{sec:gradient_flow_hypergraphs}

\com{In many applications, the state space $\X$ is associated to an underlying discrete structure.
For example, it can be the space of positive measures on the vertices of a hypergraph.
In such cases, it is more natural to consider the dynamics being generated by forces which live on the edges of the hypergraph rather than in the cotangent space of $\X$.}
\com{A} gradient flow system $(\X,\RR^*,\EE)$ can be generalized to take into account an underlying structure of a directed weighted hypergraph
\com{,which is known as} the continuity-equation format according to \cite{PeletierSchlichting2023}, and as the process-flux space formulation of gradient flow systems \cite{renger2018,renger2018flux}.
A finite oriented hypergraph $\HH = (\V,\E)$ consists of a finite vertex set $\V$ and finite edge set $\E$.
An edge $\e = (\hh(\e,.),\tt(\e,.)) \in \E$ is defined by a pair of integer valued maps $\hh(\e,.),\tt(\e,.): \V \rightarrow \mathbb{Z}$, called the head and the tail of the edge.
This gives, analogously to the graph setting, the discrete divergence operator $\div: \RE \rightarrow \RV$ and discrete gradient operator $\grad: (\RV)^* \rightarrow (\RE)^*$, which are represented by the matrices $[\div]_{\w\e} =  \tt(\e,\w) - \hh(\e,\w)$ and $[\grad]_{\e\w} = \hh(\e,\w) - \tt(\e,\w)$, respectively.
They pose a pair of negative adjoints:
\begin{align} \label{eq:adjunction}
    \div = -\grad^*.
\end{align}
It is natural to treat $\RE$ and $\left(\RE\right)^*$ as fibers of a flux bundle $\J$ and a force bundle $\F$ over $\X$, with coordinates $(\vx,\vj) \in \J$ and $(\vx,\vf) \in \F$.
Assuming that $\X$ is the space of positive measures on $\V$, the spaces $\RV$ and $\left(\RV\right)^*$ are identified with the fibers of $\T\X$ and $\T^*\X$, respectively.
The divergence and gradient operators give bundle morphisms $\div: \J \rightarrow \T\X$ and $\grad: \T^*\X \rightarrow \F$ over $\X$.

A dissipation function on a hypergraph is a function $\P^*:\F \rightarrow \R$ which satisfies the properties \ref{prop:10}-\ref{prop:30}.
It induces the Legendre duality between forces and fluxes as $\vj = \D_{\vf} \P^*(\vx,\vf)$ and $\vf = \D_{\vj} \P(\vx,\vj)$, where $\P: \J \rightarrow \R$ is the Legendre transform of $\P^*$ with respect to the second argument.
Analogously to $\RR$, the dissipation function $\P$ generalizes the Onsager reciprocity relations.
A gradient flow system on a hypergraph is the tuple $(\X,\P^*,\EE,\HH)$ with the dynamics given by the continuity equation
\begin{align}
\label{eq:dynamics_HG_1}
    \dot{\vx} &= -\div \vj, \\ 
\label{eq:dynamics_HG_2}
    \vj &= \D_{\vf} \P^*(\vx, \grad[-\D_{\vx} \EE(\vx)]).
\end{align}
The discussion regarding the gradient flow system $(\X,\RR^*,\EE)$ remains valid in this context:
\begin{align}
    \label{eq:Young_Fenchel_Psi}
    &\textrm{The pair $\P$ and $\P^*$ satisfies the inequality }\P(\vx,\vj) + \P^*(\vx,\vf) \geq \langle \vj, \vf \rangle.\\
    \label{eq:sigma_hypergraph_unperturbed}
    &\textrm{The dissipation rate is given by }\s(\vx,\vj) := \P(\vx,\vj) + \P^*(\vx,\grad[-\D_{\vx} \EE(\vx)]) \\
    &\textrm{The energy-dissipation principle holds.}
\end{align}
\begin{remark} \label{rmk:non-perturbed}
Note that, if the evolution equations (\ref{eq:dynamics_HG_1})-(\ref{eq:dynamics_HG_2}) are satisfied, then $\s(\vx,\vj) = \langle \vj, \grad[-\D_{\vx}\EE(\vx)] \rangle =  \langle -\div \vj, -\D_{\vx}\EE(\vx) \rangle =  - \dot{\EE}(\vx)$\com{,} which coincides with the dissipation rate $\s(\vx,\vv)$ of the $(\X,\RR^*,\EE)$ gradient flow system in case the gradient flow equation (\ref{eq:dynamics}) is satisfied.
Again, $\EE$ is a Lyapunov function for the dynamics.
We also remark that in the physics literature it is customary to compute the dissipation rate via the bilinear product $\langle \vj, \vf \rangle$ for a Legendre dual pair $\vf$ and $\vj$.
The definition (\ref{eq:sigma_hypergraph_unperturbed}) is more general as it accounts for possible fluctuations of the flux vector.
\end{remark}
The advantage of moving to the hypergraph setting is the possibility to describe perturbations which often arise on the edges of the hypergraph in physical contexts and which cannot be captured by $(\X,\RR^*,\EE)$ type systems.

\subsection{Perturbed gradient flow systems on hypergraphs and their energetics} \label{sec:perturbed_systems}

We describe a nonequilibrium system by a gradient flow system on a hypergraph $(\X,\P^*,\EE,\HH)$ where the driving force in Equation (\ref{eq:dynamics_HG_1}) is perturbed by an $\vx$-dependent (and possibly time-dependent) external force $\fex \in \RE$.
In other words, the perturbation is given by a (possibly time-dependent) continuous section $\fe: \X \rightarrow \F$.
This results in the perturbed evolution equation
\begin{align}
\label{eq:dynamics_HG_1_perturbed}
    \dot{\vx} &= -\div \vj, \\ 
\label{eq:dynamics_HG_2_perturbed}
    \vj &= \D_{\vf} \P^*(\vx, \grad[-\D_{\vx} \EE(\vx)] + \fex).
\end{align}
This perturbed gradient flow system is denoted by the quintuple $(\X,\P^*,\EE,\HH, \fe)$.
For any $(\vx,\vj) \in \J$, the dissipation rate  now acquires an additional contribution $\se(\vx,\vj) = \langle \vj, \fex \rangle$ due to the external force:
\begin{align}
\label{eq:sigma_ext1}
    \s(\vx,\vj) &= \P(\vx,\vj) + \P^*(\vx,\grad[-\D_{\vx} \EE(\vx)] + \fex) \\
\label{eq:sigma_ext}
    &\geq \langle \vj, \grad[-\D_{\vx} \EE(\vx)] + \fex \rangle = - \dot{\EE}(\vx) + \se(\vx,\vj),
\end{align}
where the inequality becomes an equality iff (\ref{eq:dynamics_HG_2_perturbed}) is satisfied and the last equality holds true iff (\ref{eq:dynamics_HG_1_perturbed}) is satisfied.
For the dynamics of the system, this is a drastic modification as $\EE$ is no longer a Lyapunov function and phenomena such as oscillations and multistability can occur even when $\fex$ is \com{independent of $\vx$ and }time.
The energy-dissipation balance for a differentiable curve $(\vx,\vj) : [0,T] \rightarrow \J$ with $\dot{\vx} = -\div \vj$ also acquires a perturbation in the form of the external dissipation $\DDe[(\vx,\vj)(\cdot)] := \int_0^T \se(\vx,\vj) \dt$ as
\begin{align} \label{eq:EDP_perturbed}
    \EE(\vx(T)) + \DD[(\vx,\vj)(\cdot)] \geq \EE(\vx(0)) + \DDe[(\vx,\vj)(\cdot)],
\end{align}
where the system's dissipation is $\DD[(\vx,\vj)(\cdot)] = \int_0^T \P(\vx,\vj) + \P^*(\vx, \grad[-\D_{\vx} \EE(\vx)] + \fex) \dt$.
The equality holds if and only if $\vj$ satisfies the Legendre duality (\ref{eq:dynamics_HG_2_perturbed}).
We refer to \cite{mielke2016deriving} for more details on perturbed gradient flow systems and now focus on the information\com{-}geometric aspects.

\com{
\begin{example}[Chemical reaction networks as perturbed gradient flow systems on hypergraphs]
\label{ex:CRN}

To make the theory more concrete, we show how chemical reaction network (CRN) theory can be phrased as a gradient flow system.
In Appendix \ref{app:HJB}, we discuss the connection of this formulation to large deviations theory and to the Hamilton-Jacobi formulation \cite{smith2011large,gao2023large}.

Originally, the gradient flow formulation for CRNs with quadratic dissipation functions has been used in \cite{mielke2011gradient}.
However, in \cite{mielke2014} it has been discovered that it is more natural to use $\cosh$-type dissipation as they provide a link to large deviations theory.
For an extensive overview of the relation between the gradient flow formulation CRN theory and information geometry, we refer to \cite{kobayashi2024information} and references therein.

A reversible chemical reaction network (CRN) is a hypergraph with its vertices $\chem_1,\chem_2,\dotsc,\chem_n$ corresponding to the chemical species and its edges $\reac_1,\reac_2,\dotsc,\reac_m$ corresponding to reactions.
With this choice the flux and force bundles become $\J \simeq \X \times \R^m$ and $\F \simeq \X \times (\R^m)^*$, with the respective fibers $\Jx \simeq \R^m$ and $\Fx \simeq (\R^m)^*$.
The reactions $\reac_r, r=1,\dotsc,m$ are commonly written as maps between formal linear combinations of chemicals 
\begin{align}
    \sum_{i=1}^n \tt(\reac_r,\chem_i) \chem_i \mapsto \sum_{i=1}^n \hh(\reac_r,\chem_i)\chem_i
\end{align}
with the coefficients $\hh(\reac_r,\chem_i)$ and $\tt(\reac_r,\chem_i)$ determining the hypergraph structure in accordance with Section \ref{sec:gradient_flow_hypergraphs}.
In CRN theory, the $n \times m$ matrix $\mathsf{S}$ with matrix elements
\begin{equation}
    \mathsf{S}_{ij} :=  \hh(j,i) - \tt(j,i)
\end{equation}
is called the stoichiometric matrix.
It corresponds to the operator $-\div$ in our general setting.

The state space is the space of positive concentration vectors $\vx = (x_1,x_2,\dotsc,x_n)^{\t} \in \X := \R^n_{>0}$.
The dynamics of CRNs is a classical subject, where the flux in the large volume limit for non-interacting particles is given by the law of mass action as follows:
Fix a set of reaction rate vectors $\vk^+, \vk^- \in \R_{>0}^m$ and denote the one-way fluxes at a point $\vx$ by $\vj^{0+}(\vx) = (j^{0+}_1(\vx),j^{0+}_2(\vx),\cdots,j^{0+}_m(\vx))^\t \in \R^m$ and $\vj^{0-}(\vx) = (j^{0-}_1(\vx),j^{0-}_2(\vx),\cdots,j^{0-}_m(\vx))^\t \in \R^m$, with components
\begin{align} \label{eq:j+}
    j^{0+}_r(\vx) &:= k_r^+ \prod_{i=1}^n x_i^{\tt(\reac_r,\chem_i)} \\
    \label{eq:j-}
    j^{0-}_r(\vx) &:= k_r^- \prod_{i=1}^n x_i^{\hh(\reac_r,\chem_i)}.
\end{align}
The vector of mass action net fluxes is 
\begin{equation}
\label{eq:j_0}
   \vj^0(\vx) = \vj^{0+}(\vx) - \vj^{0-}(\vx).
\end{equation}
The concentration dynamics with mass action is given by the continuity equation 
\begin{equation}
    \dot{\vx} = -\div \vj^0(\vx).
\end{equation}

This evolution equation can be recovered by a perturbed gradient flow system as follows:
The energy function is given by
\begin{equation} \label{eq:energy}
    \EE(\vx) = \sum_{i=1}^n x_i \log \frac{x_i}{x_i^*} - x_i + x_i^*,
\end{equation}
where $\vx^* \in \X$ is a fixed reference state which satisfies the detailed balancing condition
\begin{align}
    \grad \log \vx^* = \log \frac{\vkk^+}{\vkk^-}.
\end{align}
Here, $\vkk^+, \vkk^- \in \R^m_{>0}$ are reaction rate vectors which are different from $\vk^+, \vk^-$ in general, and the quotients and logarithms of vectors are taken componentwise\footnote{This condition can be equivalently written as 
\begin{equation} \label{eq:kappa_detailed}
    \k_r^+ \prod_{i=1}^n (x_i^*)^{\tt(\reac_r,\chem_i)} = \k_r^- \prod_{i=1}^n (x_i^*)^{\hh(\reac_r,\chem_i)}.
\end{equation}
\com{The reference point $\vx^*$ and the rate constants $\vkk^+, \vkk^-$ are not determined by the gradient flow system.
They must either supplied additionally based on physical considerations or chosen arbitrarily.
Their choice does, however, not affect the information-geometric constructions in Sections \ref{sec:Info_Gradient_flow} and \ref{sec:noneq_physics}.
Moreover, we refer to Remark \ref{rmk:f_b} for a natural choice of the $\vkk^+, \vkk^-$ from the point of view of information geometry.}
}.
The dissipation function $\P^*(\vx,\vf)$ has the form
\begin{align} \label{eq:CRN_dual_dissipation}
    \P^*(\vx,\vf) = 4 \sum_{r=1}^m \om \left( \cosh{\frac{f_r}{2}} - 1 \right).
\end{align}
where we have used the $\vx$-dependent frenetic activity 
\begin{align} \label{eq:CRN_omega}
    \om =\sqrt{j^{0+}_r(\vx)j^{0-}_r(\vx)}= \sqrt{k_r^+ k_r^- \prod_{i=1}^n x_i^{\tt(\reac_r,\chem_i)+ \hh(\reac_r,\chem_i)}}.
\end{align}
Following \cite{mielke2017,kobayashi2024information}, the dual dissipation potential $\P(\vx,\vj)$ can be directly computed as
\begin{align} \label{eq:CRN_primal_dissipation}
    \P(\vx,\vj) = \sum_{r=1}^m \left[ 2 j_r \arcsinh \frac{j_r}{2 \om} - 2 \sqrt{4\om^2 + j_r^2} + 4 \om \right].
\end{align}
The gradient flow system is given by (\ref{eq:dynamics_HG_1_perturbed})-(\ref{eq:dynamics_HG_2_perturbed}) with the external driving force
\begin{align} \label{eq:CRN_fex}
    \fex = \log \frac{\vk^+}{\vk^-} - \log \frac{\vkk^+}{\vkk^-}.
\end{align}
A direct calculation of $\vj^0$ via (\ref{eq:dynamics_HG_2_perturbed}), i.e., $\vj^0 = \D_{\vf} \P^*(\vx, \grad[-\D_{\vx} \EE(\vx)] + \fex)$, yields the mass action kinetics (\ref{eq:j_0}).
In Appendix \ref{app:HJB}, we discuss how the gradient flow formulation contains information on the large deviation structure of the underlying microscopic jump process away from the typical trajectory given by $\vj^0(\vx)$.

Also note that for $\fex = \vc{0}$, one obtains mass action kinetics with the rate constant vectors $\vkk^+, \vkk^-$ for the detailed balanced system and that the external driving force rescales these rate vectors by $\vk^+/\vkk^+$ and $\vk^-/\vkk^-$, respectively (the quotients of vectors are taken componentwise).
Physically, this rescaling can be thought of as the result of the presence of external chemical reservoirs.
The concentrations of the reservoirs do not directly appear in the rate equation in this case.
This can be derived in the limit of infinite reservoir size by using the limiting procedure introduced in \cite{mielke2023non}.

\begin{remark}
The energy function $\EE$ of the gradient flow system given in (\ref{eq:energy}) is the generalized Kullback-Leibler divergence between $\vx$ and $\vx^*$.
It is strictly convex in $\vx$ and therefore can be used to do information geometry on the space $\X$.
The dual coordinates are given by the Legendre transform $\D_{\vx} \EE(.)$, the Riemannian metric by the Hessian of $\EE$, etc.
This geometry has been thoroughly developed in \cite{kobayashi2021,sughiyama2021}, and has been applied to characterize growing systems \cite{sughiyama2022chemical}, finite time driving \cite{loutchko2022}, and chemical sensitivity \cite{loutchko2025information}.
However, we do not discuss this geometry in this article.
\end{remark}
\end{example}
}

\section{Information\com{-}geometric aspects of gradient flow systems} \label{sec:Info_Gradient_flow}

The setup discussed so far lends itself naturally to an information\com{-}geometric description.
We establish the basic information\com{-}geometric notions of the Bregman divergence, the associated Riemannian metric structure, and dual foliations, and comment on their physical relevance.

The information geometry takes place over a fixed point $\vx \in \X$, within the fibers $\Jx$ and $\Fx$\com{,} which we treat as a pair of dually flat affine spaces (in the coordinates $\vj$ and $\vf$, respectively) equipped with the strictly convex functions $\P(\vx,.)$ and $\P^*(\vx,.)$ and the natural bilinear pairing $\langle.,.\rangle:\Jx \times \Fx \rightarrow \R$ between dual vector spaces.
The Legendre duality between $\P(\vx,.)$ and $\P^*(\vx,.)$ induces a pair of mutually inverse diffeomorphisms $\D_{\vj}\P(\vx,.): \Jx \rightarrow \Fx$ and $\D_{\vf}\P^*(\vx,.): \Fx \rightarrow \Jx$ between the fibers.

\subsection{The Bregman divergence and large deviations theory} \label{sec:Bregman_LDT}

The Bregman divergence between two points $\vj, \vj' \in \Jx$ is given by 
\begin{align} \label{eq:def_Bregman}
    \DBJ_{\vx}[\vj \| \vj'] := \P(\vx,\vj) - \P(\vx,\vj') - \langle \vj - \vj', \D_{\vj}\P(\vx,\vj') \rangle.
\end{align}
The Bregman divergence satisfies $\DBJ_{\vx}[\vj \| \vj'] \geq 0$ and $\DBJ_{\vx}[\vj \| \vj'] = 0$ iff $\vj = \vj'$.
However, it is not symmetric in the two arguments, and therefore it is often referred to as pseudo-distance.
Analogously, for $\vf, \vf' \in \Fx$, one defines $\DBF_{\vx}[\vf \| \vf'] := \P^*(\vx,\vf) - \P^*(\vx,\vf') - \langle \D_{\vf}\P^*(\vx,\vf'), \vf - \vf' \rangle$. For a Legendre dual pair $\vf = \D_{\vj}\P(\vx,\vj)$ and $\vf' = \D_{\vj}\P(\vx,\vj')$, one verifies
\begin{align}
    \DBJ_{\vx}[\vj \| \vj'] = \DBF_{\vx}[\vf' \| \vf].
\end{align}
Moreover, the mixed representation
\begin{align}
    \DB_{\vx}[\vj \| \vf'] :&= \P(\vx,\vj) + \P^*(\vx,\vf')  - \langle \vj, \vf' \rangle \\
    \nonumber
    &= \DBJ_{\vx}[\vj \| \vj'].
\end{align}
for $\vf' = \D_{\vj}\P(\vx,\vj')$ is convenient to use.
This representation shows that the Bregman divergence quantifies the failure of the Young-Fenchel relation (\ref{eq:Young_Fenchel_Psi}) to be an equality and has an immediate physical meaning through its relation to the energetic quantities of the gradient flow system as one computes just like in (\ref{eq:sigma_ext}):
\begin{align} \label{eq:D_Bregman_perturbed}
    \DB_{\vx}[\vj \| \grad[-\D_{\vx} \EE(\vx)] + \fex] = \s(\vx,\vj) + \dot{\EE}(\vx) - \se(\vx,\vj).
\end{align}
Hence, the EDP inequality (\ref{eq:EDP_perturbed}) can be refined to yield the information\com{-}geometric EDP
\begin{multline}
\label{eq:EDP_perturbed_info_geom}
    \EE(\vx(T)) + \DD[(\vx,\vj)(\cdot)] \\
    = \EE(\vx(0)) +  \DDe[(\vx,\vj)(\cdot)] + \int_0^T \DB_{\vx}[\vj \| \grad[-\D_{\vx} \EE(\vx)] + \fex] \dt
\end{multline} 
for any differentiable curve $(\vx,\vj) : [0,T] \rightarrow \J$ with $\dot{\vx} = -\div \vj$.

Furthermore, the Bregman divergence (\ref{eq:D_Bregman_perturbed}) has a probabilistic interpretation as the \com{Lagrangian $\mathcal{L}_{\textrm{net}}(\vx,\vj)$ of the level 2.5 net flux path rate functional} of a stochastic jump process on the hypergraph in the large volume limit \cite{mielke2014,mielke2017,renger2018,renger2018flux}\com{, up to a factor of 2:
\begin{equation} \label{eq:L_net_Bregman}
     \mathcal{L}_{\textrm{net}}(\vx,\vj) = \frac{1}{2} \DB_{\vx}[\vj \| \grad[-\D_{\vx} \EE(\vx)] + \fex].
\end{equation}
More details on the large deviations interpretation are given in Appendix \ref{app:HJB}.}
In other words, it quantifies the exponential unlikeliness of the flux $\vj$ to deviate from the macroscopic flux given by (\ref{eq:dynamics_HG_2_perturbed}).
Hence, the last term in (\ref{eq:EDP_perturbed_info_geom}) is the quantifier for the exponential ``unlikeliness'' of the curve $(\vx,\vj) : [0,T] \rightarrow \J$ to be realized by a stochastic system.

\subsection{Riemannian metric structure and covariance} \label{sec:Riemannian_IG}

Continuing with this probabilistic interpretation of the Bregman divergence (\ref{eq:D_Bregman_perturbed}) as a rate function, it is a standard fact in large deviations theory that the inverse covariance matrix of the flux vector $\vj$ is given by the Hessian of the \com{Lagrangian}, which\com{, based on the relation (\ref{eq:L_net_Bregman}),} can be explicitly computed as
\begin{align} \label{eq:gJ}
   [\mathrm{Cov}_{\vx}(\vj)]^{-1} = \textcolor{black}{\frac{1}{2} \D^2_{\vj} \DB_{\vx}[\vj \| \grad[-\D_{\vx} \EE(\vx)] + \fex]} = \textcolor{black}{\frac{1}{2} \D^2_{\vj} \P(\vx, \vj)}.
\end{align}
We recognize \com{$\D^2_{\vj} \P(\vx, \vj)$} as the natural Riemannian metric on the flux space $\Jx$, known as the Fisher metric, and \com{denote} it by $\gJxj: \Tj \Jx \times \Tj \Jx \rightarrow \R$.
The Legendre duality between $\Jx$ and $\Fx$ induces an isometry with respect to the metric $\gFxf: \Tf \Fx \times \Tf \Fx \rightarrow \R$ on $\Fx$, which is given by the Hessian
\begin{align} \label{eq:gF}
    \gFxf:= \textcolor{black}{ \D^2_{\vf} \P^*(\vx,\vf) = \left[\D^2_{\vj} \P(\vx, \vj) \right]^{-1}} = \textcolor{black}{2} \mathrm{Cov}_{\vx}(\vj) 
\end{align}
in the $\vf$-coordinates.
Note that this Riemannian metric structure can be thought of as an infinitesimal version of the pseudo-distance function provided by the Bregman divergence due to the relation $\DBJ_{\vx}[\vj \| \vj + \delta \vj] = \frac{1}{2} \sum_{r,r' = 1}^{|\E|} [\gJ_{\vx,\vj}]_{rr'} \delta j_r \delta j_{r'} + \mathcal{O}(\|\delta \vj]\|^3).$

\com{
\begin{example} \label{ex:CRN_Hessian}
    For the gradient flow setting of CRN theory introduced in Example \ref{ex:CRN}, we compute explicitly
    \begin{equation} \label{eq:CRN_Hessian}
        [\mathrm{Cov}_{\vx}(\vj)]^{-1}_{rr'} = \frac{1}{2} \frac{\partial^2 \P(\vx,\vj)}{\partial j_r \partial j_{r'}} =  \frac{\delta_{rr'}}{\sqrt{j_r^2 + 4\om^2}},        
    \end{equation}
    where $\delta_{rr'} = 1$ iff $r = r'$ and $\delta_{rr'} = 0$ otherwise.
    In Appendix \ref{app:HJB}, it is shown how the minimizers $(\vj^{*+},\vj^{*-})$ of the one-way flux Lagrangian under the condition $\vj = \vj^{*+}- \vj^{*-}$ satisfy $j_r^{*+} j_r^{*-} =  \om^2$.
    Hence, the term $\sqrt{\vj^2 + 4 \vc{\omega}(\vx)^2}$ is equal to the dynamical activity $\vj^{*+} + \vj^{*-}$ based on the optimal fluxes realizing the fluctuation $\vj$ (which coincides with the dynamical activity $\vj^{0+} + \vj^{0-}$ at the mass action flux $\vj = \vj^{0+}- \vj^{0-}$).
    The inverse covariance becomes
    \begin{equation} \label{eq:CRN_Hessian}
        [\mathrm{Cov}_{\vx}(\vj)]^{-1}_{rr'} =  \frac{\delta_{rr'}}{j_r^{*+} + j_r^{*-}},      
    \end{equation}
    which is a well-known relation in the literature \cite{MaesNetocny2008Canonical,VoVuHasegawa2022UnifiedTKUR}.
\end{example}
}

\subsection{Dual foliations and moduli for perturbed gradient flow systems} \label{sec:dual_foliations}

The combination of the hypergraph structure and gradient flow structure imposes natural foliations of $\Jx$ and $\Fx$, which correspond to mixture families in information geometry.
Here, we focus on the foliation of $\Fx$ in more detail as it reveals the moduli structure of the perturbed gradient flow system, and note that duality yields the analogous statements for $\Jx$.
The two linear subspaces $\ker[\div] \subset \Jx$ and $\img[\grad] \subset \Fx$ are orthogonal with respect to the natural bilinear pairing between $\Jx$ and $\Fx$, and any linear isomorphism between the vector spaces $\Jx$ and $\Fx$ will yield an orthogonal decomposition of $\Fx$ into $\img[\grad]$ and its orthogonal complement\com{,} which is isomorphic to $\ker[\div]$.
The mixture families in information geometry generalize this situation by using Legendre duality instead of the linear isomorphism\com{,} which results in an orthogonal decomposition of $\Fx$ into $\img[\grad]$ and the Legendre transform of $\ker[\div]$ which is a non-linear space in general.
The orthogonality results from a generalized Pythagorean relation using the Bregman divergence.
We now recall the construction:
For any $\vf \in \Fx$, define the isoperturbation space of $\vf$ as
\begin{align}
    \Fpx(\vf) := \left\{ \vf' \in \Fx | \vf - \vf' \in \img[\grad] \right\}
\end{align}
and the isovelocity space of $\vf$ as
\begin{align}
    \Fox(\vf) := \left\{ \vf' \in \Fx | \D_{\vf}\P^*(\vx,\vf) - \D_{\vf}\P^*(\vx,\vf') \in \ker[\div] \right\}.
\end{align}
For any $\vj \in \Jx$, one can analogously define the isovelocity space $\Jox(\vj) := \{ \vj' \in \Jx | \vj - \vj' \in \ker[\div] \}$ and the isoperturbation space $\Jpx(\vj) := \{ \vj' \in \Jx | \D_{\vj}\P(\vx,\vj) - \D_{\vj}\P(\vx,\vj') \in \img[\grad] \}$ on the flux space.
The Legendre transformations $\D_{\vj}\P(\vx,.)$ and $\D_{\vf}\P^*(\vx,.)$ restrict to diffeomorphisms between $\Fpx(\vf)$ and $\Jpx(\vj)$ as well as between $\Fox(\vf)$ and $\Jox(\vj)$ iff $\vf$ and $\vj$ are Legendre dual.

\begin{figure}[h]
\centering
\includegraphics[width=0.95\textwidth]{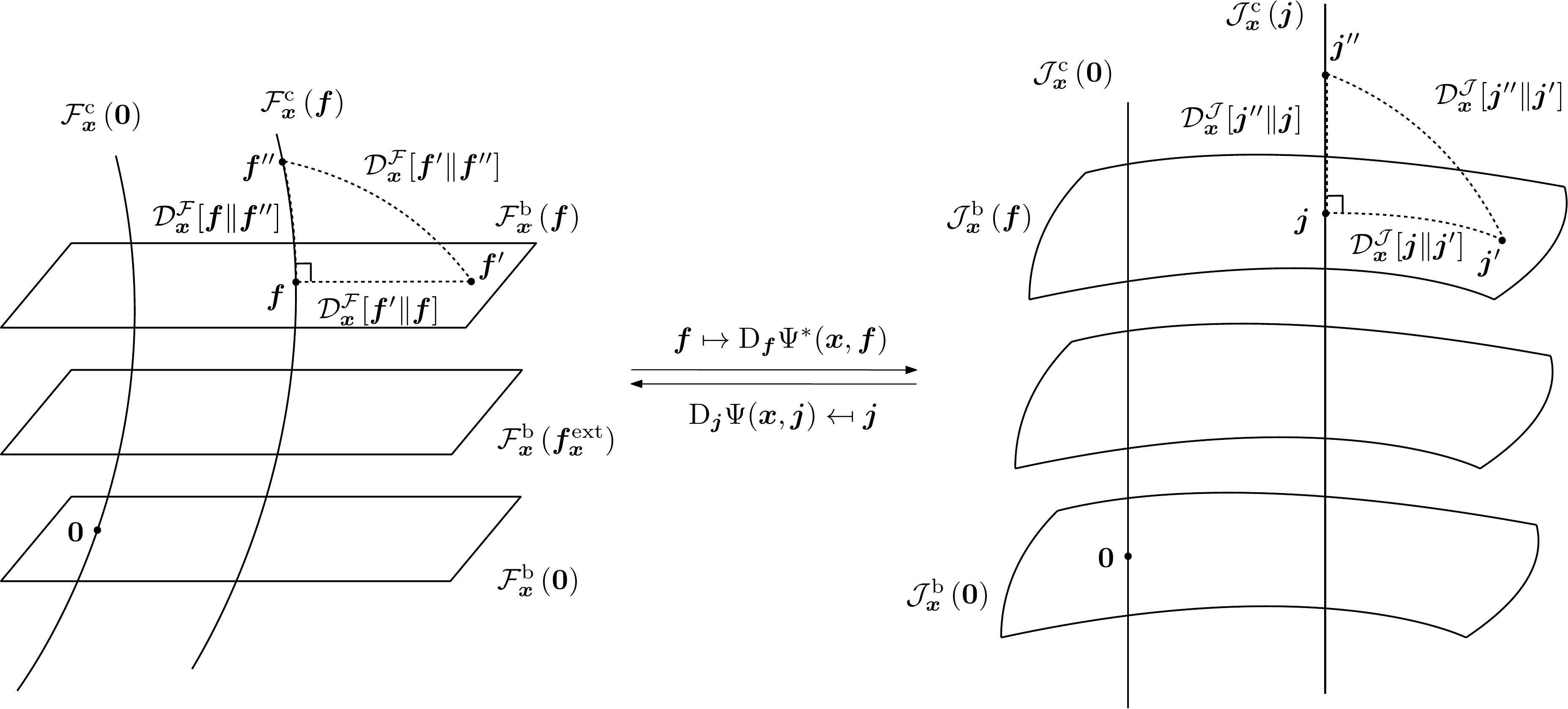}
\caption{Illustration of the dual foliations of $\Fx$ and $\Jx$ resulting from the interplay between the hypergraph structure and information geometry.
Each space $\Fox(\vf')$ acts as a base for a foliation of $\Fx$ with leaves $\{\Fpx(\vf) | \vf \in \Fox(\vf')\}$, and dually, each space $\Fpx(\vf)$ is a base space for a foliation with leaves $\{\Fox(\vf') | \vf' \in \Fpx(\vf) \}$.
The analogous situation holds for the space $\Jx$.
The spaces $\Fpx(\vf)$ and $\Jox(\vj)$ are affine while $\Fox(\vf)$ and $\Jpx(\vj)$ are curved.
The leaves intersect the base orthogonally according to the \com{Pythagorean} relation (\ref{eq:Pythagoras}), which is also illustrated in the figure.
}\label{fig1}
\end{figure}

The following generalized Pythagorean relation holds for any $\vf' \in \Fpx(\vf)$ and $\vf'' \in \Fox(\vf)$:
\begin{align} \label{eq:Pythagoras}
    \DBF_{\vx}[\vf'\| \vf''] = \DBF_{\vx}[\vf'\| \vf] + \DBF_{\vx}[\vf\| \vf''], 
\end{align}
which follows from the definition (\ref{eq:def_Bregman}) and the orthogonality condition $\langle \vf' - \vf, \D_{\vf}\P^*(\vx,\vf) - \D_{\vf}\P^*(\vx,\vf'') \rangle = 0$ between $\ker[\div]$ and $\img[\grad]$.
The generalized Pythagorean relation implies that $\vf$ is the unique intersection point\footnote{
To be precise, the relation (\ref{eq:Pythagoras}) implies that $\vf$ has the variational characterization $\vf = \argmin_{\vf' \in \Fpx(\vf)}\DBF_{\vx}[\vf'\| \vf'']$ for a fixed $\vf'' \in \Fox(\vf)$ and thus is unique.
} of $\Fpx(\vf)$ and $\Fox(\vf)$.
Thus, the space $\Fx$ is foliated by the spaces $\{\Fox(\vf') | \vf' \in \Fpx(\vf) \}$ for any $\vf \in \Fx$, i.e., any $\Fpx(\vf)$ serves as the base of the foliation.
Dually, any $\Fox(\vf')$ is a base for the foliation $\{\Fpx(\vf) | \vf \in \Fox(\vf')\}$ of $\Fx$.
The Legendre transform gives the analogous foliations $\{\Jox(\vj') | \vj' \in \Jpx(\vj) \}$ and $\{\Jpx(\vj) | \vj \in \Jox(\vj') \}$ of $\Jx$\com{,} which can be obtained from the ones of $\Fx$ via the Legendre transformation.
This geometry is illustrated in Figure \ref{fig1}.

This geometrical setup can be used to construct the moduli space for the perturbed gradient flow system (\ref{eq:dynamics_HG_1_perturbed})-(\ref{eq:dynamics_HG_2_perturbed}) as follows:
First, we require the following definitions.

\begin{definition} \label{def:f_eq}
For a given gradient flow system on a hypergraph $(\X,\P^*,\EE,\HH)$, two continuous perturbation fields $\fe, \fet: \X \rightarrow \F$ are equivalent if and only if
\begin{align}
    \Fpx\left(\fex\right) = \Fpx\left(\fext \right)
\end{align}
for all $\vx \in \X$.
In other words, if and only if the perturbations $\fe $ and $\fet$ differ by a section $\vc{h}_{\vx} := \fex - \fext$, which can be written as a discrete gradient, i.e., $\vc{h}_{\vx} \in \img[\grad]$.
\end{definition}

\begin{remark} \label{rmk:coarse_moduli}
    This definition of equivalence is rather coarse as we do not require the existence of a potential $\EEt : \X \rightarrow \R$ such that $\vc{h}_{\vx} = \grad[-\D_{\vx} \EEt(\vx)]$.
    \com{Such additional requirements for restricted classes of potentials $\EEt$ lead} to finer moduli problems which we do not address here.
    \com{However, such finer moduli problems can be rather intricate, and have not been studied in the literature to the best of our knowledge.}
\end{remark}

\begin{definition} \label{def:moduli}
    A {\it moduli space} for continuous perturbation fields $\fe: \X \rightarrow \F$ of a gradient flow system $(\X,\P^*,\EE,\HH)$ is a set that is in one-to-one correspondence with the equivalence classes (according to Definition \ref{def:f_eq}) of continuous perturbation fields.
\end{definition}

Definition \ref{def:f_eq} states that, for a given $\vx \in \X$, the equivalence classes of perturbation fields $\fex$ are given by the leaves of the foliation $\{\Fpx(\vf) | \vf \in \Fox(\vf')\}$, which we call the {\it local} moduli space for the moduli problem \ref{def:moduli}.
In particular, choosing the base space $\Fox(\vc{0})$ for the foliation and using the Legendre duality between $\vf = \vc{0}$ and $\vj = \vc{0}$, one sees that $\Jox(\vc{0})$, which is the linear subspace $\ker[\div] \subset \Jx$, is a local moduli space.
This gives the solution to the (global) moduli problem:
\begin{theorem} \label{thm:moduli}
    The moduli space for continuous perturbation fields $\fe: \X \rightarrow \F$ of a gradient flow system $(\X,\P^*,\EE,\HH)$ is given by the set of continuous sections $\{ s: \X \rightarrow \K \}$, where $\K$ denotes the subbundle of $\J$ with the fibers $\K_{\vx} = \ker[\div] \subset \Jx$.
\end{theorem}
This is a generalized reformulation of the Schnakenberg theory for the decomposition of nonequilibrium forces at steady states into cycle components \cite{schnakenberg1976}.
Note that the classical Schnakenberg theory treats the case of $\vx$-independent perturbation fields $\fe$ where one obtains one fiber $\K_{\vx} = \ker[\div]$ as the corresponding moduli space:
\com{
\begin{example}[Schnakenberg theory as a moduli problem]
    For the CRN setup from Example \ref{ex:CRN}, we make the connection to classical Schnakenberg theory more explicit.
    It is worth noting that the theory in \cite{schnakenberg1976} only treats graphs, whereas CRN theory requires the hypergraph setting.
    After choosing an integer basis $\{\vz_1,\vz_2,\cdots,\vz_q\}$ of $\ker[\div] \subset \Jx$, where $q$ is the dimension of $\ker[\div]$, the cycle affinities $(A_1,A_2,\cdots,A_q)$ are given by
    \begin{equation}
        A_i = \langle \vz_i, \vf^0 \rangle
    \end{equation}
    with $\vf^0 = \log \frac{\vj^{0+}(\vx)}{\vj^{0-}(\vx)} = \grad[-\D_{\vx} \EE(\vx)] + \fex$.
    This explicitly gives
    \begin{equation}
        A_i = \sum_{r=1}^m c_{ir} \log \frac{k^+_r}{k^-_r}.
    \end{equation}
    Schnakenberg theory states that the system admits a detailed-balanced equilibrium with rate constants $(\vk^+,\vk^-)$ if and only if all cycle affinities vanish.
    In our setting this is equivalent to the equilibrium condition $\vf^0 \in \img[\grad].$

    More abstractly, we can say that the natural bilinear pairing $\Jx \times \Fx \rightarrow \R$ restricts to a degenerate pairing $\ker[\div] \times \Fx \rightarrow \R$ and descends to the nondegenerate pairing
    \begin{equation}
        \ker[\div] \times \bigsfrac{\Fx}{\img[\grad]} \rightarrow \R.
    \end{equation}
    Under this pairing, the cycle affinities are precisely the components of the equivalence class \([\vf^0]\in \bigsfrac{\Fx}{\img[\grad]}\) with respect to the chosen cycle basis, i.e., $A_i =  \langle \vz_i,[\vf^0]\rangle$, and this definition is independent of the representative of $[\vf^0]$.  
    Using $\K_{\vx} = \ker[\div] \simeq \bigsfrac{\Fx}{\img[\grad]}$, we see that $\K_{\vx}$ is a moduli space for cycle affinities.
    The globalized version of Schnakenberg theory stated in Theorem \ref{thm:moduli} is the natural setting for $\vx$-dependent rate constants $(\vk^+,\vk^-)$.
\end{example}
}

In the next section, we present some typical physical applications which utilize information and differential geometrical techniques.
We show how information geometry can be used to streamline known results from statistical physics and to derive new ones.

\section{Nonequilibrium physics via the information geometry to gradient flow systems} \label{sec:noneq_physics}

Within the recent decades, nonequilibrium physics has seen the discovery of modern principles such as fluctuation theorems \cite{seifert2012stochastic,korbel2021stochastic,martins2025brief}, thermodynamic uncertainty relations \cite{barato2015thermodynamic,horowitz2020thermodynamic,falasco2020unifying,koyuk2022thermodynamic,van2023thermodynamic,kamijima2023thermodynamic,ray2023thermodynamic}, speed limits \cite{vo2020unified,yoshimura2021thermodynamic,nagayama2025infinite}, and entropy production rate decompositions \cite{dechant2022geometric,kobayashi2022geometry,yoshimura2023housekeeping}.

We give \com{three} example applications which have been developed in the context of chemical reaction network (CRN) theory \cite{kobayashi2022geometry,loutchko2023geometry} \com{but are now} phrased more generally in the setting of perturbed gradient flow systems.
\com{This does not require} the particular details from CRN theory for the most part.
\com{More precisely, we cover dissipation rate decompositions in Section \ref{sec:decomposition}, and thermodynamic uncertainty relations and speed limits in Section \ref{sec:Riemannian_application}.}
\com{Moreover, as in Section \ref{sec:decomposition} we are able to treat the flux $\vj$ and the force $\vf$ as independent variables, i.e., not necessarily Legendre duals, in the definition of the dissipation rate}\footnote{
\com{
This is common for gradient flow systems as the flux is allowed to fluctuate.
Meanwhile, the physical force should be set to $\grad[-\D_{\vx} \EE(\vx)] + \fex$.}}.

\subsection{Dissipation rate decomposition} \label{sec:decomposition}

The decomposition of the dissipation rate into separate contributions from cycle and boundary components of the force and flux vectors has a long\com{-}standing history in the theory of \com{nonequilibrium} physics, where it is known as entropy production rate decomposition \cite{schnakenberg1976,oono1998steady,hatano2001steady,komatsu2011entropy,maes2014nonequilibrium,rao2016nonequilibrium,dechant2022geometric,yoshimura2023housekeeping}.
As pointed out in Footnote \ref{footnote:dissipation}, the concept of dissipation rate is more general and hence we keep using this terminology.
Such decompositions have also been treated in the gradient flow literature \cite{renger2021orthogonality,gao2022revisit}.
We refer to \cite{kobayashi2022geometry} for an extended discussion.

The cycle components are naturally identified as fluxes $\vjc$ in the linear subspace $\Jox(\vc{0}) = \ker[\div]$ whereas the boundaries are forces $\vfb$ in the linear space $\Fpx(\vc{0}) = \img[\grad]$.
Recall from Eq. (\ref{eq:sigma_ext1}) the definition of the dissipation rate as
\begin{align} \label{eq:sigma_ext3}
    \sigma(\vx,\vj) = \P(\vx,\vj) + \P^*(\vx,\grad[-\D_{\vx} \EE(\vx)] + \fex). 
\end{align}
The great advantage of this gradient flow systems representation is the additive splitting of the contributions from the flux $\vj$ and the force $\grad[-\D_{\vx} \EE(\vx)] + \fex$ through the dual dissipation potentials.
This allows to consider \com{dissipation} rate decomposition where the flux is not given by Eq. (\ref{eq:dynamics_HG_2_perturbed}), i.e., by $\vj = \D_{\vf} \P^*(\vx, \grad[-\D_{\vx} \EE(\vx)] + \fex)$, but is allowed to fluctuate - this situation is not treated within the classical approaches.

We recall the decomposition for a pair of quadratic dissipation potentials $\P(\vx,\vj) = \frac{1}{2}\vj^\t \L(\vx) \vj$ and $\P^*(\vx,\vj) = \frac{1}{2}\vf^\t \L(\vx)^{-1} \vf$ (where $\L(\vx)$ is a symmetric positive definite matrix) \cite{dechant2022geometric}.
Usually, this decomposition is derived under the condition that $\vj$ is the Legendre dual of $\vf$.
However, here, we do not require this condition.
The quadratic dissipation potentials naturally equip the spaces with $\L(\vx)$, respectively $\L(\vx)^{-1}$-weighted bilinear product structures\footnote{
The bilinear product $\langle .,. \rangle_{\L(\vx)}$ on $\Jx$ is defined as $\langle \vc{u}, \vc{v} \rangle_{\L(\vx)} := \sum_{i,j=1}^{|\E|} u_i \L(\vx)_{ij} v_j$ for $\vc{u}, \vc{v} \in \Jx$, and the \com{bilinear} product $\langle .,. \rangle_{\L(\vx)^{-1}}$ is defined analogously on $\Fx$.
}
and the respective vector norms $\| . \|_{\L(\vx)}$ and $\| . \|_{\L(\vx)^{-1}}$.
The dissipation rate is then given by the sum of squared norms $\sigma(\vx,\vj) = \frac{1}{2} \| \vj \|_{\L(\vx)}^2 + \frac{1}{2} \| \vf \|_{\L(\vx)^{-1}}^2$, and a natural decomposition is obtained by letting $\vjc$ be the $\langle.,.\rangle_{\L(\vx)}$-orthogonal projection of $\vj$ to $\Jox(\vc{0})$, and $\vfb$ be the $\langle.,.\rangle_{\L(\vx)^{-1}}$-orthogonal projection of $\vf$ to $\Fpx(\vc{0})$.
Let $\vjb = \vj - \vjc$ and $\vfc = \vf - \vfb$ be the respective orthogonal complements.
Note that we have $\D_{\vj}\P(\vx,\vjb) \in \Fpx(\vc{0})$ as well as $\D_{\vf}\P^*(\vx,\vfc) \in \Jox(\vc{0})$.
This yields the desired decomposition into the housekeeping and excess dissipation rates $\sigma^{\textrm{hk}}_{\textrm{IG}}(\vx,\vj)$ and $\sigma^{\textrm{ex}}_{\textrm{IG}}(\vx,\vj)$:
\begin{align} \label{eq:MN_decomposition}
    \sigma(\vx,\vj) = \underbrace{\frac{1}{2} \left[ \|  \vjc \|_{\L(\vx)}^2 +  \|  \vfc \|_{\L(\vx)^{-1}}^2  \right]}_{\sigma^{\textrm{hk}}_{\textrm{MN}}(\vx,\vj)}    
    + \underbrace{\frac{1}{2}  \left[ \|  \vjb \|_{\L(\vx)}^2 +  \|  \vfb \|_{\L(\vx)^{-1}}^2  \right] }_{\sigma^{\textrm{ex}}_{\textrm{MN}}(\vx,\vj)},
\end{align}
which separates the effect of the boundary and cycle components.
When $\vj$ and $\vf$ are Legendre dual, one verifies the equalities $\|  \vjc \|_{\L(\vx)}^2 = \|  \vfc \|_{\L(\vx)^{-1}}^2$ and $ \|  \vjb \|_{\L(\vx)}^2 = \|  \vfb \|_{\L(\vx)^{-1}}^2$.

\begin{figure}[h]
\centering
\includegraphics[width=0.90\textwidth]{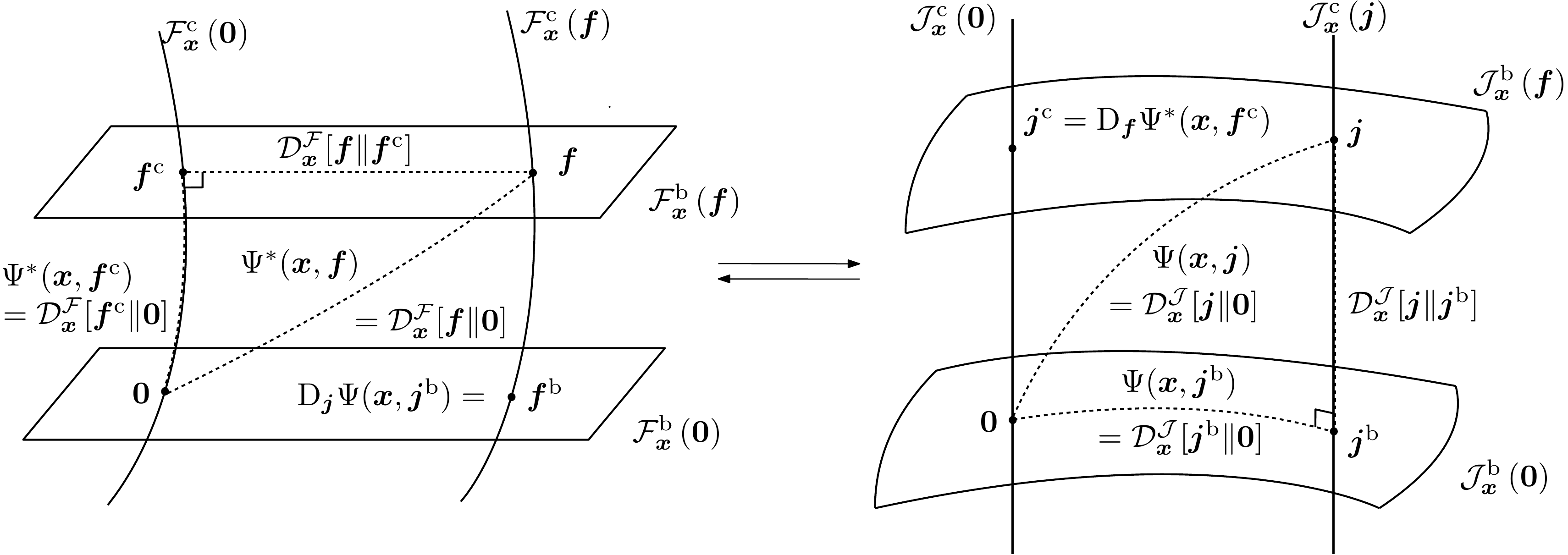}
\caption{Illustration of the geometry underlying the dissipation rate decomposition (\ref{eq:IG_decomposition}). On $\Fx$, the composition is \com{achieved} via the cycle force $\vfc = \Fpx(\vf) \cap \Fox(\vc{0})$ and the \com{corresponding} orthogonal decomposition of $\P^*(\vx,\vf)$, and dually, on $\Jx$ via the boundary flux $\vjb = \Jpx(\vc{0}) \cap \Jox(\vj)$.
Note that one can also define directly $\vfb := \Fpx(\vc{0}) \cap \Fox(\vf)$ to get the Legendre dual of $\vjb$, and analogously for $\vjc := \Jpx(\vj) \cap \Jox(\vc{0})$.}\label{fig2}
\end{figure}

We now discuss the generalization of this decomposition to the case of arbitrary dissipation potentials following \cite{kobayashi2022geometry}.
The orthogonal \com{splitting} of the respective \com{dissipation} potentials uses the dual orthogonal foliations introduced in Section \ref{sec:dual_foliations}, which generalize the orthogonal decompositions from linear algebra.
The following discussion applies for arbitrary $\vf$, i.e, not \com{necessarily} of the form $\vf = \grad[-\D_{\vx} \EE(\vx)] + \fex$, and arbitrary $\vj$.
Let $\vfc \in \Fx$ be the \com{unique} intersection point of $\Fpx(\vf)$ and $\Fox(\vc{0})$.
The dissipation potential for arbitrary $\vf' \in \Fx$ can be represented as $\P^*(\vx,\vf') = \DBF_{\vx}[\vf'\| \vc{0}] $, and thus the generalized Pythagorean relation (\ref{eq:Pythagoras}) yields 
\begin{align} \label{eq:Psi_decomp1}
\nonumber
    \P^*(\vx,\vf) &= \DBF_{\vx}[\vf \| \vc{0}] = \DBF_{\vx}[\vf \| \vfc] + \DBF_{\vx}[\vfc \| \vc{0}] \\
    &= \DBF_{\vx}[\vf \| \vfc] + \P^*(\vx,\vfc).
\end{align}
Analogously, let $\vjb \in \Jx$ be the \com{unique} intersection point of $ \Jox(\vj)$ and $\Jpx(\vc{0})$.
The dual Pythagorean relation yields the decomposition 
\begin{align} \label{eq:Psi_decomp2}
    \P(\vx,\vj) =  \DBJ_{\vx}[\vj \| \vjb] + \P(\vx,\vjb).
\end{align}
This generalizes the decomposition (\ref{eq:MN_decomposition}) as
\begin{align} \label{eq:IG_decomposition}
    \sigma(\vx,\vj) = 
    \underbrace{\left[  \P^*(\vx,\vfc) + \DBJ_{\vx}[\vj \| \vjb]  \right]}_{\sigma^{\textrm{hk}}_{\textrm{IG}}(\vx,\vj)}    
    + \underbrace{\left[ \P(\vx,\vjb) + \DBF_{\vx}[\vf \| \vfc] \right] }_{\sigma^{\textrm{ex}}_{\textrm{IG}}(\vx,\vj)}.
\end{align}
This is the decomposition proposed in \cite{kobayashi2022geometry}.
The geometry of this decomposition is illustrated in Fig. \ref{fig2}.

We briefly discuss the interpretation of information\com{-}geometric decomposition \ref{eq:IG_decomposition} for the case that the gradient flow equations (\ref{eq:dynamics_HG_1_perturbed})-(\ref{eq:dynamics_HG_2_perturbed}) hold true.
If the system is unperturbed, then $\Fpx(\vf) = \Fpx(\vc{0})$ and $\vfc = \Fpx(\vc{0}) \cap \Fox(\vc{0}) = \vc{0}$ hold.
This yields $\vfb = \vf$ and, by Legendre duality, $\vjb = \vj$.
Thus, only the excess dissipation rate survives and we have $\sigma(\vx,\vj) = \sigma^{\textrm{ex}}_{\textrm{IG}}(\vx,\vj)$ for the unperturbed case.

At a steady state $\xss$, which is defined by $\xssd = \vc{0}$, the continuity equation $\dot{\vx} = -\div \vj$ implies that $\vj \in \Jox(\vc{0})$ and thus $\vjb = \Jox(\vc{0}) \cap \Jpx(\vc{0}) = \vc{0}$.
Moreover, the Legendre duality between $\vj$ and $\vf$ yields $\vf \in \Fox(\vc{0})$ and therefore $\vfc = \Fpx(\vf) \cap \Fox(\vc{0}) = \vf$.
This shows that the excess dissipation rate $\sigma^{\textrm{ex}}_{\textrm{IG}}(\xss,\vj)$ vanishes and we obtain $\sigma(\xss,\vj) = \sigma^{\textrm{hk}}_{\textrm{IG}}(\xss,\vj)$.

Hence, for a perturbed gradient flow system, the excess term generalizes the dissipation away from equilibrium to the dissipation away from the steady state, and relates it to the boundary components of fluxes and forces.
The housekeeping term quantifies the dissipation due to the nonequilibrium nature of the system, i.e., due to the presence of the perturbation term, and relates it to the cycle components.

\com{
\begin{remark} \label{rmk:f_b}
     With respect to the arbitrary choice of detailed balancing rate constants $\vkk^+, \vkk^-$ introduced in Example \ref{ex:CRN}, the point $\vfb := \Fpx(\vc{0}) \cap \Fox(\vf)$ provides us with natural candidates for $\vkk^+, \vkk^-$ from the point of view of information geometry.
     Because $\vfb$ lies in $\Fpx(\vc{0}) = \img[\grad] \subset \Fx$, and because $\D_{\vx} \EE(\vx) = \sum_{i=1}^n x_i \log \frac{x_i}{x_i^*}$ with $\EE(\vx)$ defined in Equation (\ref{eq:energy}) spans the whole $\T^*_{\vx} \X$ for fixed $\vx$ and variable $\vx^* \in \X$, we find that a reference point $\vx^*$ exists such that
     \begin{equation}
         \vfb = \grad [-  \D_{\vx} \EE(\vx)].
     \end{equation}
     Then, the rate constants $\vkk^+, \vkk^-$ are given by Equation (\ref{eq:kappa_detailed}) up to a multiplicative factor scaling each pair $\kappa^+_r$ and $\kappa^-_r$.
     However, there is no necessity for this construction to be compatible at different points of the base manifold $\X$, i.e., such vectors $\vkk^+, \vkk^-$ arising would be $\vx$-dependent.
     This incompatibility would introduce an $\vx$-dependence in the external force $\fex$ which can be avoided in CRN theory if the rate constants $\vk^+, \vk^-$ themselves are $\vx$-independent.
\end{remark}
}

\subsection{\com{Riemannian} geometry of thermodynamic uncertainty relations and speed limits} \label{sec:Riemannian_application}

In this section, we present a coherent information\com{-}geometric toolbox to deal with thermodynamic \com{uncertainty} relations and thermodynamic speed limits.

\subsubsection{Thermodynamic uncertainty relations} \label{sec:TUR}

Thermodynamic \com{uncertainty} relations (TURs) have been extensively studied in the physics literature for over a decade, and various formulations have been obtained based on explicit models \cite{barato2015thermodynamic,horowitz2020thermodynamic,koyuk2022thermodynamic,kamijima2023thermodynamic,ray2023thermodynamic}.
Throughout this section, we assume that the gradient flow equations (\ref{eq:dynamics_HG_1_perturbed})-(\ref{eq:dynamics_HG_2_perturbed}) hold true, set $\vf_{\vx} =  \grad[-\D_{\vx} \EE(\vx)] + \fex$, and denote the corresponding flux by $\vj := \D_{\vf} \P^*(\vx, \vf_{\vx})$.
\com{In this case, the dissipation rate is given by
\begin{equation}
    \sigma(\vx,\vj) = \P(\vx,\vj) + \P^*(\vx,\vf_{\vx}) = \langle \vj, \vf_{\vx} \rangle = \langle \vj, \D_{\vj} \P(\vx, \vj) \rangle
\end{equation}
which follows from the definition (\ref{eq:sigma_ext1}) and the Legendre duality between $\vj$ and $\vf_{\vx}$ implying the representation as a bilinear product and $\vf_{\vx} = \D_{\vj} \P(\vx, \vj)$.
}

The main idea of a TUR is to bound the relative precision $\px(\vj)$ of a flux, defined as
\begin{align} \label{eq:precision}
    \px(\vj):= \vj^\t [\mathrm{Cov}_{\vx}(\vj)]^{-1} \vj = \textcolor{black}{\frac{1}{2} \vj^\t \left[ \D^2_{\vj} \P(\vx, \vj) \right] \vj} j_r j_{r'} = \textcolor{black}{\frac{1}{2}} \|\vj\|_{\gJxj}^2,
\end{align}
through the dissipation rate $\sigma(\vx,\vj)$.
Here, using the notation from Section \ref{sec:Riemannian_IG} we denote by $\|.\|_{\gJxj}$ the norm associated to the inner product $\gJxj$ on $\Tj\Jx$\footnote{
Note that this notation requires to treat $\vj$ as an element of the tangent space $\Tj\Jx$ by making the canonical identification $\Jx \simeq \Tj\Jx$ of a vector space with its tangent space at a point via $e_r \mapsto \partial_r$.
The same is required in the integral representation of the dissipation rate in formula (\ref{eq:int_EPR}).
}.
\com{We now derive} the \com{following} integral representation of the dissipation rate
\begin{align} \label{eq:int_EPR}
    \sigma(\vx,\vj) = \int_0^1 \| \vj \|^2_{\mathsf{g}^{\mathcal{J}}_{\vx,\tau\vj}} \dd \tau.
\end{align}
\com{To this end, consider the line $\gamma:[0,1] \rightarrow \Jx$ given by $\gamma(\tau):= \tau \vj$.
Along the line, we have
\begin{align}
    \frac{\dd}{\dd \tau} \D_{\vj} \P(\vx, \gamma(\tau)) = \left[ \D^2_{\vj} \P(\vx, \gamma(\tau)) \right] \vj,
\end{align}
which, using the property \ref{prop:30}, i.e., $\D_{\vj} \P(\vx, \vc{0}) = \vc{0}$, integrates to
\begin{equation}
    \D_{\vj} \P(\vx, \vj) = \int_0^1 \left[\D^2_{\vj} \P(\vx, \gamma(\tau))\right] \vj \dd \tau.
\end{equation}
Multiplication with $\vj^\t$ from the left, and recognizing $\D^2_{\vj} \P(\vx, \gamma(\tau))$ as the inner product $\mathsf{g}^{\mathcal{J}}_{\vx,\tau\vj}$ on $\T_{\tau \vj}\Jx$ gives the desired relation (\ref{eq:int_EPR}).
We remark that the integral representation (\ref{eq:int_EPR}) can also be derived from integral representation of the Bregman divergence introduced in \cite{AyAmari2015CanonicalDivergences} after symmetrization (see also \cite{amari2016information}, Theorem 4.2.).
}

With the definition (\ref{eq:precision}), the formula (\ref{eq:int_EPR}) gives the following general result on TUR:
\begin{lemma} \label{lemma:TUR}
If $\| \vj \|_{\mathsf{g}^{\mathcal{J}}_{\vx,\tau\vj}} \geq \| \vj \|_{\mathsf{g}^{\mathcal{J}}_{\vx,\vj}}$ holds for all $\tau \in [0,1]$, then the relative precision of the flux is bounded by the dissipation rate as
\begin{align} \label{eq:TUR_lemma}
    \frac{1}{2} \sigma(\vx,\vj) \geq \px(\vj).
\end{align}
\end{lemma}
\com{
\begin{example}
    A direct calculation shows the condition $\| \vj \|_{\mathsf{g}^{\mathcal{J}}_{\vx,\tau\vj}} \geq \| \vj \|_{\mathsf{g}^{\mathcal{J}}_{\vx,\vj}}$ for $\tau \in [0,1]$ to hold true for the CRN setting introduced in Example \ref{ex:CRN}.
    By using the explicit expression for the Hessian metric ${\mathsf{g}^{\mathcal{J}}_{\vx,\tau\vj}}$ from Example \ref{ex:CRN_Hessian}, cf. Equation (\ref{eq:CRN_Hessian}), we compute explicitly
    \begin{align}
    \| \vj \|_{\mathsf{g}^{\mathcal{J}}_{\vx,\tau\vj}} = \sum_{r=1}^{m} \frac{2j_r^2}{\sqrt{(\tau j_r)^2 + 4\om^2}} \geq\sum_{r=1}^{m} \frac{2j_r^2}{\sqrt{ j_r^2 + 4\om^2}} =  \| \vj \|_{\mathsf{g}^{\mathcal{J}}_{\vx,\vj}}.
\end{align}
This holds true for all $\tau \in [0,1]$.
\end{example}
}
The same argument \com{as in the above example} also goes through more generally for $\cosh$-type gradient flow systems \cite{PeletierSchlichting2023}.

In Section \ref{sec:infinitesimal_decomp}, we show how this implies generalizations of some known TURs.
We leave it as an open question to establish \com{the} condition \com{$\| \vj \|_{\mathsf{g}^{\mathcal{J}}_{\vx,\tau\vj}} \geq \| \vj \|_{\mathsf{g}^{\mathcal{J}}_{\vx,\vj}}$ for $\tau \in [0,1]$} under more general assumptions.

\subsubsection{Speed limits}

In addition to the TUR, the integral representation (\ref{eq:int_EPR}) allows to \com{obtain} a clean geometrization of thermodynamic speed limits.
Speed limits relate the time $T$ to transition between two points $\vx_0$ and $\vx_T$ to the dissipation and have been extensively studied in the physics literature \cite{vo2020unified,yoshimura2021thermodynamic,nagayama2025infinite}.

Let $(\vx,\vjx) : [0,T] \rightarrow \J$ be a curve which obeys the gradient flow equations (\ref{eq:dynamics_HG_1_perturbed})-(\ref{eq:dynamics_HG_2_perturbed}).
Recall from Section \ref{sec:perturbed_systems} that the total dissipation is given by $\DD[(\vx,\vj_{\vx})(\cdot)] = \int_0^T \sigma(\vx,\vj_{\vx}) \dt$\com{,} which yields, via the formula (\ref{eq:int_EPR}):
\begin{align}
    \DD[(\vx,\vj_{\vx})(\cdot)] = \int_0^T \int_0^1 \| \vjx \|^2_{\mathsf{g}^{\mathcal{J}}_{\vx,\tau\vjx}} \dd \tau \dd t.
\end{align}
By using the Cauchy-Schwarz inequality one obtains 
\begin{align}
    \DD[(\vx,\vj_{\vx})(\cdot)] \geq \frac{1}{T} \left[ \int_0^T \int_0^1 \| \vjx \|_{\mathsf{g}^{\mathcal{J}}_{\vx,\tau\vjx}} \dd \tau \dd t \right]^2.
\end{align}
Then, by defining the {\it thermodynamic area} $\A_{\vx_0:\vx_T}$ between two points $\vx_0$ and $\vx_T$ via
\begin{align}
    \A_{\vx_0:\vx_T} := \inf_{\substack{
    (\vx,\vjx) : [0,T] \rightarrow \J \\
    \vx(0) = \vx_0 \\
    \vx(T) = \vx_T \\
    \dot{\vx} = - \div \vj }}  \int_0^T \int_0^1 \| \vjx \|_{\mathsf{g}^{\mathcal{J}}_{\vx,\tau\vjx}} \dd \tau \dd t,
\end{align}
we obtain the information\com{-}geometric speed limit:
\begin{align}
    T  \geq \frac{\A_{\vx_0:\vx_T}^2}{\DD[(\vx,\vj_{\vx})(\cdot)]}.
\end{align}
The speed limit relates the transition time between $\vx_0$ and $\vx_T$ to the inverse dissipation via the thermodynamic area.
This area term is mathematically similar to the concept of thermodynamic length \cite{salamon1983thermodynamic,crooks2007measuring,sivak2012thermodynamic,mandal2016analysis,loutchko2022}.
However, the area describes the dissipation in a genuine \com{nonequilibrium} situation whereas the length treats slowly driven equilibrium systems.

\subsubsection{Infinitesimal decompositions} \label{sec:infinitesimal_decomp}

We end this subsection by combining the discussion of information\com{-}geometric dissipation rate decompositions presented in Section \ref{sec:decomposition} with TURs and speed limits\com{,} which are best treated with the infinitesimal methods of Riemannian geometry presented in this subsection.

The relative precision of the flux vector $\vj$ splits into the relative precision of the velocity $\dot{\vx} = -\div \vj $ and the one of the cycle flux component $\vjc \in \Jox(\vj)$ as follows:
For any flux vector $\vj \in \Jx$, one verifies that the spaces $\Jox(\vj)$ and $\Jpx(\vj)$ intersect $\gJxj$-orthogonally at $\vj$, i.e., that the tangent spaces $\Tj \Jox(\vj)$ and $\Tj \Jpx(\vj)$ are \com{orthogonal} with respect to the inner product on $\Tj \Jx$ defined by the Riemannian metric $\gJxj$.
Moreover, the tangent spaces $\Tj \Jpx(\vj)$ and $\Tj \Jox(\vj)$ span the whole space $\Tj \Jx$.
We write
\begin{align} \label{eq:orthogonal_tangent_decomposition}
    \Tj \Jx = \Tj \Jpx(\vj) \oplus_{\gJxj} \Tj \Jox(\vj)
\end{align}
for this decomposition.
Denote the respective $\gJxj$-orthogonal projections onto \com{$\Tj \Jpx(\vj)$} by $\ppx$ and onto \com{$\Tj \Jox(\vj)$} by $\pox$, respectively.
One verifies that the boundary component of the flux \com{gives} the velocity, i.e.,
\begin{align}
    -\div \ppx(\vj) = \dot{\vx}
\end{align}
and that its orthogonal complement $\vjc := \pox(\vj)$ lies in $\ker[\div]$ and is thus a cycle component.
The orthogonal decomposition (\ref{eq:orthogonal_tangent_decomposition}) leads to the decomposition of the relative precision as
\begin{align} \label{eq:precision_Riemannian_decomposition}
    \px(\vj)= \frac{1}{2} \|\vj\|_{\gJxj}^2 = \underbrace{ \frac{1}{2} \ \| \pox(\vj) \|_{\gJxj}^2}_{\px(\vjc)} + \underbrace{ \frac{1}{2} \ \| \ppx(\vj) \|_{\gJxj}^2}_{\px(\dot{\vx})}
\end{align}
where $\px(\dot{\vx})$ is the relative precision of the velocity and $\px(\vjc)$ is the relative precision of the cycle component $\vjc$.
Then, \com{combining} this with Lemma \ref{lemma:TUR} (which requires the gradient flow system to satisfy the assumption of the lemma) yields the inequality $\frac{1}{2}   \sigma(\vx,\vj) \geq \px(\dot{\vx})$\com{,} which sharpens and generalizes the TUR derived in \cite{yoshimura2021thermodynamic}, as well as the inequality $\frac{1}{2} \sigma(\vx,\vj) \geq \px(\vjc)$\com{,} which recovers the TUR from \cite{loutchko2023geometry}.

Finally, we remark that the decomposition (\ref{eq:precision_Riemannian_decomposition}) can be used in the integral formula (\ref{eq:int_EPR}) to obtain a Riemannian geometric variation of the dissipation rate decomposition (\ref{eq:IG_decomposition}) as
\begin{align} \label{eq:RG_decomposition}
    \sigma(\vx,\vj) = 
    \underbrace{ \int_0^1 \| \pox(\vj) \|^2_{\mathsf{g}^{\mathcal{J}}_{\vx,\tau\vj}} \dd \tau}_{\sigma^{\textrm{hk}}_{\textrm{RG}}(\vx,\vj)}    
    + \underbrace{ \int_0^1 \| \ppx(\vj)  \|^2_{\mathsf{g}^{\mathcal{J}}_{\vx,\tau\vj}} \dd \tau}_{\sigma^{\textrm{ex}}_{\textrm{RG}}(\vx,\vj)}
\end{align}
As discussed in last paragraph of Section \ref{sec:decomposition}, the term $\sigma^{\textrm{ex}}_{\textrm{RG}}(\vx,\vj)$ captures the dissipation due \com{to} the relaxation dynamics whereas the term $\sigma^{\textrm{hk}}_{\textrm{RG}}(\vx,\vj)$ is related to the nonequilibrium nature of the system due to the perturbation.
One can \com{also} tighten the TURs above \cite{loutchko2023geometry} to obtain
\begin{align}
\label{eq:TUR_split1}
    \frac{1}{2} \sigma^{\textrm{ex}}_{\textrm{RG}}(\vx,\vj) &\geq \px(\dot{\vx}) \\
\label{eq:TUR_split2}
    \frac{1}{2} \sigma^{\textrm{hk}}_{\textrm{RG}}(\vx,\vj) &\geq \px(\vjc). 
\end{align}
Note that\com{,} by summing both the inequalities (\ref{eq:TUR_split1}) and (\ref{eq:TUR_split2}) one recovers the original TUR (\ref{eq:TUR_lemma}).
From the geometrical point of view, according to (\ref{eq:orthogonal_tangent_decomposition}) this is the only possible orthogonal splitting of the TUR because any other decomposition - when applied to the integral expression (\ref{eq:int_EPR}) would necessarily introduce mixture terms.

\section{Future perspectives} \label{sec:discussion}

In this article, we have shown how the gradient flow system formalism can be \com{naturally} transferred to an information\com{-}geometric setting:
The Bregman divergence is tightly connected to the dissipation, the Fisher metric to \com{covariance}, and the dual orthogonal foliations provide natural cycle-boundary decompositions and moduli spaces.
We have then shown how to use the geometric formulation to derive dissipation rate decompositions, TURs, and speed limits.
It is worth stressing that these modern \com{principles} discovered in nonequilibrium physics are derived without any specific physical \com{model} but only require the gradient flow structures.
As such, the results presented here make such principles applicable to any system which can be described by a gradient flow evolution equation.
In \com{particular}, it will be interesting to discuss such principles within the gradient flow formulation of machine learning \cite{liero2025evolution}.

This work should be thought of as \com{a} first foray into the information geometry of gradient flow systems, and we hope to stimulate further developments.
We briefly mention some possible directions:

We have extensively used the finite dimensionality of the state space and of the vector bundles on that space.
The information geometry on infinite dimensional state spaces is well-developed \cite{ay2017information}, and gradient flow systems on infinite dimensional state space are well-known.
\com{Although} merging these two formalisms is technically demanding, it will be interesting to develop this theory.

There are coordinate-free formulations of both gradient flow systems \cite{Mielke2025EulerianLieDerivativesGENERIC} and of information geometry \cite{amari2000methods}, and it will be rewarding to formulate the nonequilibrium \com{principles} in this setting.
Connected to this is the study of dissipation potentials which are not strictly convex, as the corresponding information \com{geometry} for singular models has been recently developed \cite{nakajima2021dually}.

Most importantly, we have not touched the bread and butter of gradient flow systems which are thermodynamically consistent and mathematically rigorous limit\com{ing} procedures (discrete to \com{continuous}, time-scale separations).
Such limits should have nice geometric \com{counterparts}, and should allow to study the \com{nonequilibrium} physical principles in such situations.

\bmhead{Acknowledgements}
We thank Matthias Liero for fruitful discussions and for educating us on technical details as well as on the intuition of gradient flow systems.

\section*{Declarations}

\bmhead{Funding}
This research is supported by JST (JPMJCR2011, JPMJCR25Q2) and JSPS (24K00542,  25H01365). 
The second author is financially supported by JST SPRING, Grant No. JPMJSP2108.

\bmhead{Conflict of interest}
The authors have no conflict of interest to declare.

\bmhead{Code and data availability}
Data sharing is not applicable to this article as no new data was created or analyzed.

\begin{appendices}

\section{Explicit differential-geometric setup}
\label{app:diff_geom}
Differential geometry is a central ingredient in gradient flow systems as the vector fields which govern the evolution of the system live on the tangent bundle of the state space, whereas the forces which drive the evolution live on the cotangent bundle.
Usually, the state space $\X$ admits a global coordinate system which has a physical interpretation and is therefore most commonly used for explicit constructions \cite{mielke2014,liero2017microscopic}.
However, there are also coordinate-free formulations of the framework \cite{Mielke2025EulerianLieDerivativesGENERIC}.
In this section\com{,} we make the notations from Section \ref{sec:gradient_flow} explicit under the assumption that $\X$ is a finite-dimensional real manifold without boundary and admits a global coordinate system, i.e., it is an open submanifold of some $\R^n$.\\

We denote the coordinates of a point $\vx \in \X$ by the vector $\vx = (x_1,x_2,\dotsc,x_n)^{\t} \in \R^n$.
At each point $\vx$, the tangent space $\T_{\vx}\X$ is spanned by the tangent vectors $\left\{\left. \frac{\partial}{\partial x_i} \right|_{\vx} \right\}_{i=1}^n$ in the directions of the $x_i$ at the point $\vx$.
The tangent space is isomorphic to $\R^n$.
With a global coordinate system on $\X$, the tangent spaces glue to the tangent bundle
\begin{equation} \label{eq:bundle_TX}
    \T\X \simeq \X \times \R^n
\end{equation}
with the global coordinates $(\vx,\vv) \in \T\X$, where
$\vv = (v_1,v_2,\dotsc,v_n)^{\t} \in \R^n$ is given by
\begin{equation}
    \vv = \sum_{i=1}^n v_i \left. \frac{\partial}{\partial x_i} \right|_{\vx}.
\end{equation}
We denote the dual vector space of $\R^n$, i.e., the space of linear functions from $\R^n$ to $\R$, by $(\R^n)^*$.
To define the cotangent bundle, let $\R^n$ be spanned by $\left\{\left. \frac{\partial}{\partial x_i} \right|_{\vx} \right\}_{i=1}^n$, and define the dual basis
$\left\{\left. \dd x_i \right|_{\vx} \right\}_{i=1}^n \subset (\R^n)^*$ via the the relation
\begin{equation} \label{eq:bilinear}
    \dd x_i \left(  \frac{\partial}{\partial x_j} \right) = 
    \begin{cases}
1, \textrm{ if } i = j\\
0, \textrm{ else}
\end{cases}.
\end{equation}
The cotangent space $\T_{\vx}^*\X$ at $\vx \in \X$ is identified with $(\R^n)^*$ via the basis $\left\{\left. \dd x_i \right|_{\vx} \right\}_{i=1}^n$ and globally glued to the cotangent bundle
\begin{equation} \label{eq:bundle_T*X}
    \T^*\X \simeq \X \times (\R^n)^*
\end{equation}
with coordinates $(\vx,\vm) \in \T^*\X$, whereby $\vm = (\mu_1,\mu_2,\dotsc,\mu_n)^{\t} \in (\R^n)^*$ is given by
\begin{equation}
    \vm = \sum_{i=1}^n \mu_i \left. \dd x_i \right|_{\vx}.
\end{equation}
The bilinear pairing between $\R^n$ and $(\R^n)^*$ gives the bilinear pairing
\begin{equation}
    \langle.,.\rangle: \T_{\vx}\X \times \T_{\vx}^*\X \rightarrow \R
\end{equation}
between tangent and cotangent vectors $\vv$ and $\vm$ over a point $\vx$.
Using the coordinate systems above, together with (\ref{eq:bilinear}), one obtains
\begin{equation}
    \langle \vv , \vm \rangle = \left\langle \sum_{i=1}^n v_i \left. \frac{\partial}{\partial x_i} \right|_{\vx}, \sum_{i=1}^n \mu_i \left. \dd x_i \right|_{\vx} \right\rangle = \sum_{i=1}^n v_i \mu_i.
\end{equation}

A function $\EE: \X \rightarrow \R$ gives rise to a differential form $\left(\vx,\D_{\vx} \EE(x)\right) \in \T^*\X$ via exterior differentiation, which is explicitly given by
\begin{equation}
    \D_{\vx} \EE(x) = \sum_{i=1}^n \frac{\partial \EE(x)}{\partial x_i} \left. \dd x_i \right|_{\vx}.
\end{equation}

Moving to the setup in Sections \ref{sec:gradient_flow_hypergraphs} and \ref{sec:perturbed_systems}, for the hypergraph $\HH = (\V,\E)$, we label the vertices $\V \simeq \{1,2,\dotsc,n\}$ and edges $\E \simeq \{1,2,\dotsc,m\}$.
This identifies $\RV \simeq \R^n \simeq \T_{\vx}\X$ and $\left(\RV\right)^* \simeq (\R^n)^* \simeq \T_{\vx}^*\X$, as well as $\RE \simeq \R^m$ and $\left(\RE\right)^* \simeq (\R^m)^*$.
Define the trivial bundles
\begin{align}
\label{eq:bundle_J}
    \J &:= \X \times \R^m \\
\label{eq:bundle_F}
    \F &:= \X \times (\R^m)^*
\end{align}
with coordinates $(\vx,\vj) \in \J$ and $(\vx,\vf) \in \F$.
The fibers $\Jx \simeq \R^m$ and $\Fx \simeq (\R^m)^*$ are equipped with coordinates $\vj = (j_1,j_2,\dotsc,j_m)^{\t} \in \R^m$ and $\vf = (f_1,f_2,\dotsc,f_m)^{\t} \in (\R^m)^*$ chosen such that the natural bilinear pairing
\begin{equation}
    \langle.,.\rangle: \Jx \times \Fx \rightarrow \R
\end{equation}
is given by
\begin{equation}
    \langle \vj, \vf \rangle = \sum_{r=1}^m j_r f_r.
\end{equation}
Furthermore, define the $n \times m$ stoichiometric matrix $\mathsf{S}$ as
\begin{equation*}
    \mathsf{S}_{ij} :=  \hh(j,i) - \tt(j,i).
\end{equation*}
Then the discrete divergence operator $\div: \RE \rightarrow \RV$ and discrete gradient operator $\grad: (\RV)^* \rightarrow (\RE)^*$ correspond to $\div =-S$ and $\grad = S^\t$.
The adjunction between $-\div$ and $\grad$ in Equation (\ref{eq:adjunction}) corresponds to the transposition of the $\mathsf{S}$ matrix.
With the trivial bundle structure (\ref{eq:bundle_TX}), (\ref{eq:bundle_T*X}), (\ref{eq:bundle_J})-(\ref{eq:bundle_F}), and the coordinate systems for the fibers discussed above, this gives the maps on the fibers as
\begin{align}
    \Jx \simeq \mathbb R^m
    &\xrightarrow{-\div = S}
    \T_{\vx}\X \simeq \R^n,
    &\quad \vj &\mapsto \vv = S\vj, \\
    \T_{\vx}^*\X \simeq (\mathbb R^n)^*
    &\xrightarrow{\grad = S^\t}
    \Fx \simeq (\mathbb R^m)^*,
    &\quad \vm &\mapsto \vf = S^\t\vm .
\end{align}

\section{Connection to large deviations theory}
\label{app:HJB}
For the interested reader, we outline the connection between the gradient-flow formalism used in this manuscript and the large-deviation theory of Markov jump processes and chemical reaction networks.
We refer to \cite{FengKurtz2006} for a thorough introduction to the general theory, to \cite{smith2011large,gao2023large} for the particular aspects of the Hamilton-Jacobi approach to CRN theory, and to \cite{liero2017microscopic} for establishing the connection to gradient flow systems.

We fix the notation as in Example \ref{ex:CRN} introducing CRN theory, with one-way mass action fluxes $\vj^{0+}(\vx)$ and $\vj^{0-}(\vx)$ at a point $\vx$ given by Equations (\ref{eq:j+})-(\ref{eq:j-}) and the mass action net flux $\vj^0(\vx) = \vj^{0+}(\vx) - \vj^{0-}(\vx)$.

To account for deviations from the typical one-way fluxes in the sense of large deviations theory, we introduce the bundle $\widetilde{\J}$ of one-way fluxes over $\X$, which is given by
\begin{equation}
    \widetilde{\J} \simeq \X \times \R^{m} \times \R^m,
\end{equation}
with coordinates $(\vx,\vj^+,\vj^-) \in \widetilde{\J}$.

For a trajectory $(\vx,\vj^+,\vj^-):[0,T] \rightarrow \widetilde{\J}$ such that the continuity equation
\begin{equation}
    \dot{\vx} = - \div [\vj^+ - \vj^-]
\end{equation}
holds, the level 2.5 path rate functional is given by
\begin{equation}
    \mathcal{I}[(\vx,\vj^+,\vj^-)(\cdot)] = \int_0^T \mathcal{L}(\vx,\vj^+,\vj^-) \dd t.
\end{equation}
The integrand is the Lagrangian $\mathcal{L}(\vx,\vj^+,\vj^-)$ which is explicitly given by the sum of relative entropy costs for the respective Poisson processes:
\begin{align}
  \mathcal{L}(\vx,\vj^+,\vj^-) = \sum_{r=1}^m \left[ j^+_r \log \frac{j^+_r}{j^{0+}_{r}(\vx)} - j^+_r + j^{0+}_{r}(\vx) \right] \\
  + \sum_{r=1}^m \left[ j^-_r \log \frac{j^-_r}{j^{0-}_{r}(\vx)} - j^-_r + j^{0-}_{r}(\vx)  \right].
\end{align}
When passing to net fluxes, the contracted Lagrangian
\begin{equation} \label{eq:L_explicit}
    \mathcal{L}_{\textrm{net}}(\vx,\vj) := \inf_{\substack{
    (\vj^+,\vj^-)\\
    \vj = \vj^+ -\vj^- }} \mathcal{L}(\vx,\vj^+,\vj^-)
\end{equation}
for the path rate functional on the level of net fluxes can be explicitly computed as
\begin{align} \label{eq:L_net_explicit}
    \mathcal{L}_{\textrm{net}}(\vx,\vj) = \sum_{r=1}^m \left[ j_r \arcsinh \frac{j_r}{2 \omega_r(\vx)} - \sqrt{ j_r^2 + 4\omega_r(\vx)^2} \right. \\  
    \left. + 2 \omega_r(\vx)   \cosh{\frac{f_r^0(\vx)}{2}} - \frac{1}{2}j_r f_r^0(\vx)\right].
\end{align}
Here,
\begin{equation} \label{eq:app_w}
    \vc{\omega}(\vx) := \sqrt{\vj^{0+}(\vx) \vj^{0-}(\vx)}
\end{equation}
is the frenetic activity vector as in (\ref{eq:CRN_omega}) and $\vf^0(\vx) = (f^0_1(\vx),f^0_2(\vx),\cdots,f^0_m(\vx))^\t $ is the vector of physical forces given by
\begin{equation}
    \vf^0(\vx) := \log \frac{\vj^{0+}(\vx)}{\vj^{0-}(\vx)},
\end{equation}
whereby all arithmetic operations on vectors are performed componentwise.
The derivation of (\ref{eq:L_net_explicit}) is a standard calculation, which we now sketch for the sake of completeness:
Write the one-way flux Lagrangian $\mathcal{L}(\vx,\vj^+,\vj^-)$ as
\begin{equation}
    \mathcal{L}(\vx,\vj^+,\vj^-) = \DKL[\vj^+ \| \vj^{0+}] + \DKL[\vj^- \| \vj^{0-}]
\end{equation}
and introduce the Lagrange multipliers $\vc{\lambda} = (\lambda_1,\lambda_2,\cdots,\lambda_m)^\t \in \R^m$ for the constraint $\vj = \vj^+ - \vj^-$.
Consider the augmented objective function
\begin{equation}
    \mathcal{G}(\vx,\vj^+,\vj^-) =  \DKL[\vj^+ \| \vj^{0+}] + \DKL[\vj^- \| \vj^{0-}] - \vc{\lambda}^\t(\vj^+ - \vj^- - \vj)
\end{equation}
whose stationarity conditions read
\begin{align}
    \frac{\partial \mathcal{G}(\vx,\vj^+,\vj^-)}{\partial j_r^+} &= \log \frac{j_r^+}{j_r^{0+}} - \lambda_r = 0 \\
    \frac{\partial \mathcal{G}(\vx,\vj^+,\vj^-)}{\partial j_r^-} &= \log \frac{j_r^-}{j_r^{0-}} + \lambda_r = 0.
\end{align}
This gives the relation between the optimizers $(\vj^{*+},\vj^{*-})$ and the mass action fluxes $(\vj^{0+},\vj^{0-})$ as $\vj^{*+} = \vj^{0+} \textrm{e}^{\vc{\lambda}}$ and $\vj^{*-} = \vj^{0-} \textrm{e}^{-\vc{\lambda}}$ which characterizes the optimizers by the two equations
\begin{align}
\label{eq:j_optimizers}
    \vj^{*+} - \vj^{*-} &= \vj \\
\label{eq:j_frenetic}
    \vj^{*+} \vj^{*-} &= \vj^{0+} \vj^{0-}.
\end{align}
Using the frenetic activity from Equation (\ref{eq:app_w}), the equations are solved by
\begin{align} \label{eq:optimizers}
    \vj^{*\pm} = \frac{\sqrt{\vj^2 + 4\vc{\omega}(\vx)^2} \pm \vj}{2}.
\end{align}
Substituting the optimizers into Equation (\ref{eq:L_explicit}) gives and using (\ref{eq:j_optimizers}) now leads to the desired relation $\mathcal{L}(\vx,\vj^{*+},\vj^{*-}) = \mathcal{L}_{\textrm{net}}(\vx,\vj)$.

The energy $\EE: \X \rightarrow \R$ enters through the detailed balancing condition which requires $\vf^0 (\vx) = \grad[-\D_{\vx} \EE(\vx)]$.
This integrability condition is the classical Wegscheider condition for the existence of detailed balanced equilibria \cite{HornJackson1972}.
If it can be satisfied, then there exists an energy function of the form (\ref{eq:energy}).
Otherwise, due to the nonintegrability, an additional term $\fex$ remains, and the logarithm of the flux ratios is written as
\begin{equation} \label{eq:force}
   \vf^0 (\vx) = \log \frac{\vj^{0+}(\vx)}{\vj^{0-}(\vx)} = \grad[-\D_{\vx} \EE(\vx)] + \fex,
\end{equation}
with $\fex$ given by (\ref{eq:CRN_fex}) and $\EE(\vx)$ by the relative entropy (\ref{eq:energy}).

Using the dissipation potentials $\P(\vx,\vj)$ and $\P^*(\vx,\vf)$ given by Equations (\ref{eq:CRN_primal_dissipation}) and (\ref{eq:CRN_dual_dissipation}), the net flux Lagrangian can be written as
\begin{equation} \label{eq:L_net}
    \mathcal{L}_{\textrm{net}}(\vx,\vj) = \frac{1}{2} \left[ \P(\vx,\vj) + \P^*(\vx,\vf^0(\vx)) - \langle \vj, \vf^0(\vx) \rangle \right].
\end{equation}
Employing the Bregman divergence from Section \ref{sec:Bregman_LDT}, together with Equation (\ref{eq:force}), the Lagrangian has the information\com{-}geometric representation
\begin{equation} \label{eq:L_net_D}
     \mathcal{L}_{\textrm{net}}(\vx,\vj) = \frac{1}{2} \DB_{\vx}[\vj \| \grad[-\D_{\vx} \EE(\vx)] + \fex].
\end{equation}
If the continuity equation $\dot{\vx} = -\div \vj$ is satisfied, then, using the adjunction (\ref{eq:adjunction}), the bilinear product $\langle \vj, \vf^0(\vx) \rangle$ becomes
\begin{equation}
    \langle \vj, \vf^0(\vx) \rangle = \langle -\div \vj, -\D_{\vx} \EE(\vx) \rangle + \langle \vj, \fex  \rangle = - \dot{\EE}(\vx) + \se(\vx,\vj).
\end{equation}
Plugging this into equation (\ref{eq:L_net}), we obtain
\begin{equation} \label{eq:L_net2}
    2 \mathcal{L}_{\textrm{net}}(\vx,\vj) =  \P(\vx,\vj) + \P^*(\vx,\vf^0(\vx)) + \dot{\EE}(\vx) - \se(\vx,\vj)
\end{equation}
Based on the nonnegativity of the Lagrangian, the integration of the right hand side of (\ref{eq:L_net2}) yields the energy dissipation relation (\ref{eq:EDP_perturbed}) and forces the definition of the dissipation rate given in Equation (\ref{eq:sigma_ext1}).
This highlights the probabilistic nature of the definition of dissipation in gradient flow systems with microscopic origin \cite{liero2017microscopic}. 

We close this section with some remarks:

\begin{remark}
    In \cite{liero2017microscopic,kobayashi2022geometry,kobayashi2024information} and related papers, a different scaling convention has been used:
    In those papers, the force $\vc{\xi}$ and the dual dissipation potential $\psi^*(\vx,\vc{\xi})$ are scaled by a factor $\frac{1}{2}$ relative to the convention used in this paper, i.e., $\vc{\xi} = \frac{1}{2} \vf$, and $\psi^*(\vx,\vc{\xi}) = 2 \sum_{r} \om  [ \cosh \xi_r -1 ]$.
    This gives the same flux $\vj$ as in this paper, and gets rid of the factor 2 in the relations (\ref{eq:L_net}) and (\ref{eq:L_net_D}).
    However, because the physical force corresponds to the $\vf$-variable and not the $\vc{\xi}$-variable, the factor 2 must be reintroduced in the dissipation rate in order to relate it to physics in the cited papers.
    Because the dissipation rate plays a central in our physical application, we have chosen to use the scaling based on the physical force variable here.
\end{remark}

\begin{remark}
    We elaborate on the connection to the Hamiltonian formalism for CRNs \cite{smith2011large,gao2023large}:
    For one-way fluxes, the Hamiltonian $\mathcal{H}(\vx,\vt^+,\vt^-)$ is the Legendre dual of $\mathcal{L}(\vx,\vj^+,\vj^-)$ with respect to the $(\vj^+,\vj^-)$ variables, and is given by
    \begin{equation}
        \mathcal{H}(\vx,\vt^+,\vt^-) = \sum_{r=1}^m \left[j^{0+}_{r}(\vx) (\textrm{e}^{\theta^+_r} -1) + j^{0-}_{r}(\vx) (\textrm{e}^{\theta^-_r} -1)  \right].
    \end{equation}
    For the net flux Lagrangian $\mathcal{L}_{\textrm{net}}(\vx,\vj)$, the Legendre dual with respect to $\vj$ is given by
    \begin{equation}
        \mathcal{H}_{\textrm{net}}(\vx,\vt) = \sum_{r=1}^m \left[j^{0+}_{r}(\vx) (\textrm{e}^{\theta_r} -1) + j^{0-}_{r}(\vx) (\textrm{e}^{-\theta_r} -1)  \right].
    \end{equation}
    This can be rewritten as
    \begin{equation}
        \mathcal{H}_{\textrm{net}}(\vx,\vt) = 2\sum_{r=1}^m \left(\omega_r(\vx) \left[ \cosh \left( \theta_r + \frac{f_r^0(\vx)}{2} \right) - \cosh \frac{f_r^0(\vx)}{2} \right]  \right).
    \end{equation}
    This form clarifies the relation between the net flux Hamiltonian and the dissipation potential $\P^*(\vx,\vf)$ given in Equation (\ref{eq:CRN_dual_dissipation}) as
    \begin{equation}
        \mathcal{H}_{\textrm{net}}(\vx,\vt) = \frac{1}{2} \left[ \P^*(\vx,\vf^0(\vx) + 2 \vt) - \P^*(\vx,\vf^0(\vx))   \right].
    \end{equation}
    This shift enables the properties \ref{prop:10}, \ref{prop:20}, and \ref{prop:30} to hold which are expected from far-from equilibrium Onsager relations.
\end{remark}

\begin{remark}
We briefly explain the reason for the contraction from the one-way-flux formulation to the net-flux formulation. 
The thermodynamic interpretation of the force $\vf$ relies on the detailed-balance structure: 
the ratio of forward and backward rates can be represented in terms of the gradient of a driving energy, which leads to the unperturbed gradient flow formulation \cite{liero2017microscopic}.
In contrast, the variables $\vt^+$ and $\vt^-$ are exponential tilting parameters conjugate to the one-way empirical fluxes. 
They have a probabilistic interpretation through the moment-generating function of jump counts, but they do not have a direct thermodynamic interpretation. 
\end{remark}

\end{appendices}




\bibliography{bibliography}

\end{document}